\documentclass[12pt]{article}  
\usepackage{graphicx}

\usepackage{amsmath,amsthm,amssymb,bm}

\newcommand{\B}{Ba\v zant}  

 \newcommand{\bc}{\begin{center}}
 \newcommand{\ec}{\end{center}}
                   \newcommand{\bfr}{\begin{flushright}}
                   \newcommand{\efr}{\end{flushright}}
   
     \newcommand{\be}{\begin{enumerate}}
     \newcommand{\ee}{\end{enumerate}}
        \newcommand{\bi}{\begin{itemize}}
        \newcommand{\ei}{\end{itemize}}
            \newcommand{\bd}{\begin{description}}
            \newcommand{\ed}{\end{description}}
                \newcommand{\beq}{\begin{equation}}
                \newcommand{\eeq}{\end{equation}}
                  \newcommand{\bea}{\begin{eqnarray}}
                  \newcommand{\eea}{\end{eqnarray}}

      \newcommand{\bfi}{\begin{figure}}
      \newcommand{\efi}{\end{figure}}
\newcommand{\bay}{\begin{array}{l}}
\newcommand{\eay}{\end{array}}
            \newcommand{\dd}{\mbox{d}}

    \newcommand{\pa}{\partial}

    \newcommand{\la}{\lambda}
    
    \newcommand{\al}{\alpha}
    \newcommand{\sig}{\sigma}
    
    \newcommand{\tht}{\theta}  

\newcommand{\nn}{\nonumber}        
\newcommand{\cPr}{\mathbb{P}} 

\begin{document}
    
\thispagestyle{empty}
\hspace*{1mm}  \vspace*{-0mm}
\noindent {\footnotesize {{\em
			~~~
			\vskip 1.5in
			\begin{center}
				{\Large {\bf  Fishnet Statistics for Strength Scaling of \\[2 mm]
						Nacreous Imbricated Laminar Materials}}\\[20mm]
				
				{\large {\sc Wen Luo and Zden\v ek P. Ba\v zant}}
				\\[1.5in]
				{\sf SEGIM Report No. 17-05/fi}
				\\[2.2in]
				Department of Civil and Environmental Engineering
				\\ Northwestern University
				\\ Evanston, Illinois 60208, USA
				\\[1in]  {\bf May 30, 2017} 
			\end{center}
			
			\clearpage   \pagestyle{plain} \setcounter{page}{1}
			
\em \normalsize

\begin{center}
 {\Large {\sf  
      Fishnet Statistics for Strength Scaling of \\[1.5 mm] Nacreous 
      Imbricated Lamellar Materials
        }} \\[7mm]  {\large {\sf
      Wen Luo\footnote{
Graduate Research Assistant, Northwestern University} and
      Zden\v ek P. Ba\v zant\footnote{
McCormick Institute Professor and W.P. Murphy Professor of Civil
and Mechanical Engineering and Materials Science, Northwestern University, 2145 Sheridan Road, CEE/A135, Evanston, Illinois 60208; corresponding author, z-bazant@northwestern.edu.}
 }}\\
\end{center} \vskip 5mm   \baselineskip 13.8pt

\noindent {\bf Abstract:}\, {\sf  
        Similar to nacre or brick-and-mortar structures, imbricated lamellar structures are widely found in natural and man-made materials and are of interest for biomimetics. These structures are known to be rather insensitive to defects and to have a very high fracture toughness. Their deterministic behavior has been intensely studied, but statistical studies have been rare and is so far there is no undisputed theoretical basis for the probability distribution of strength of nacre. This paper presents a numerical and theoretical study of the PDF of strength and of the corresponding statistical size effect. After a reasonable simplifications of the shear bonds, an axially loaded lamellar shell is statistically modelled as a square fishnet pulled diagonally. A finite element (FE) model is developed and used in Monte Carlo simulations of strength. An analytical model for failure probability of the fishnet is developed and matched to the computed statistical histograms of strength for various sizes. It appears that, with increasing size, the pdf of strength slowly transits from Gaussian to Weilbull distribution but the transition is different from that previously obtained at Northwestern for quasibrittle materials of random heterogeneous mesostructure.
    }
    \tableofcontents
    
    \section{\large Introduction}
    

    Nacre-like imbricated (staggered) structures (see Fig.~\ref{fig:nacre}) are commonly seen in both nature and man-made materials, and are of biomimetic interest. The mechanical merits of their characteristic 'brick-and-mortar' structure such as high fracture toughness and insensitivity of flaws have been well established. For exammple, Wang et al. \cite{WangSuo01}
showed that the nano-asperities on the platelets of hydrated nacres are the main cause of high resistance during the sliding of platelets and the robust mechanical behavior on the macro-scale. Gao et al. \cite{Gao03} found that the stress concentrations at flaws and the damage in nacre are significantly mitigated on the nano-scale. Shao et al. \cite{Shao12} demonstrated that the fracture toughness of nacre is greatly enhanced by the crack bridging effects in the staggered microstructure.
   \begin{figure}[!h]
       \centering
       \includegraphics[width=0.3\textwidth]{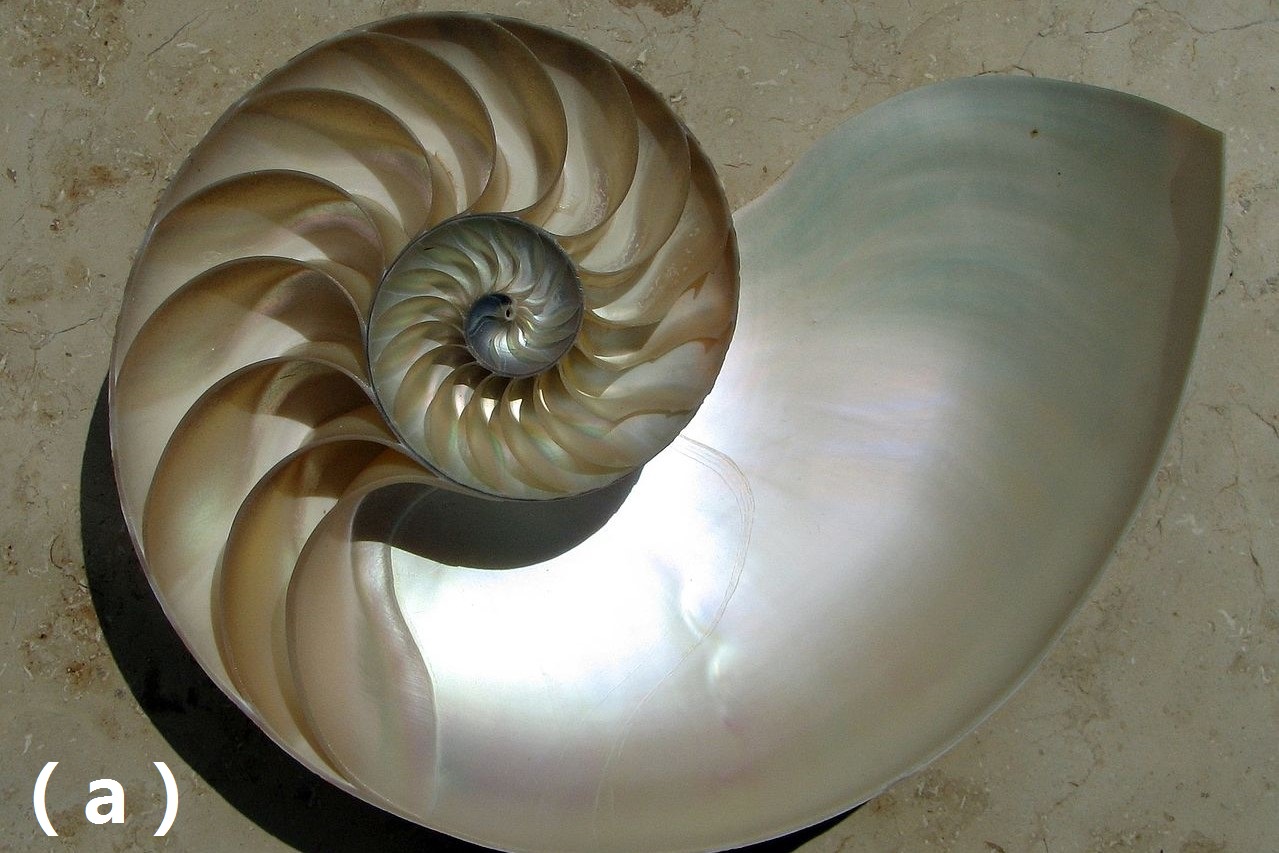}
       ~
       \includegraphics[width=0.315\textwidth]{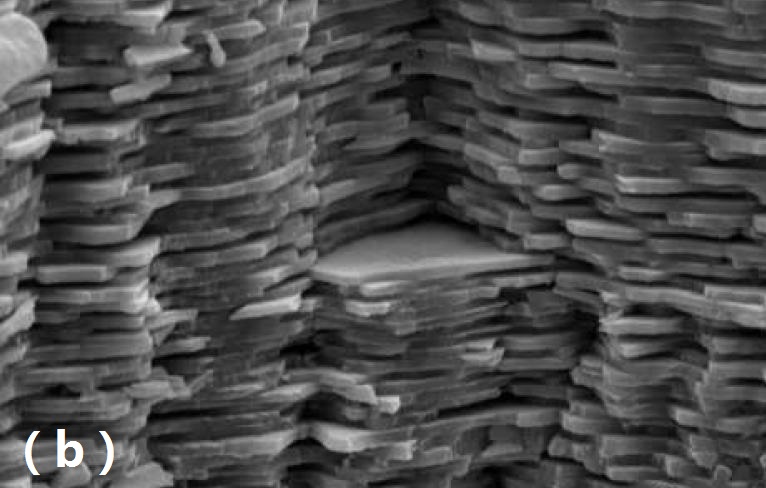}
       ~
       \includegraphics[width=0.3\textwidth]{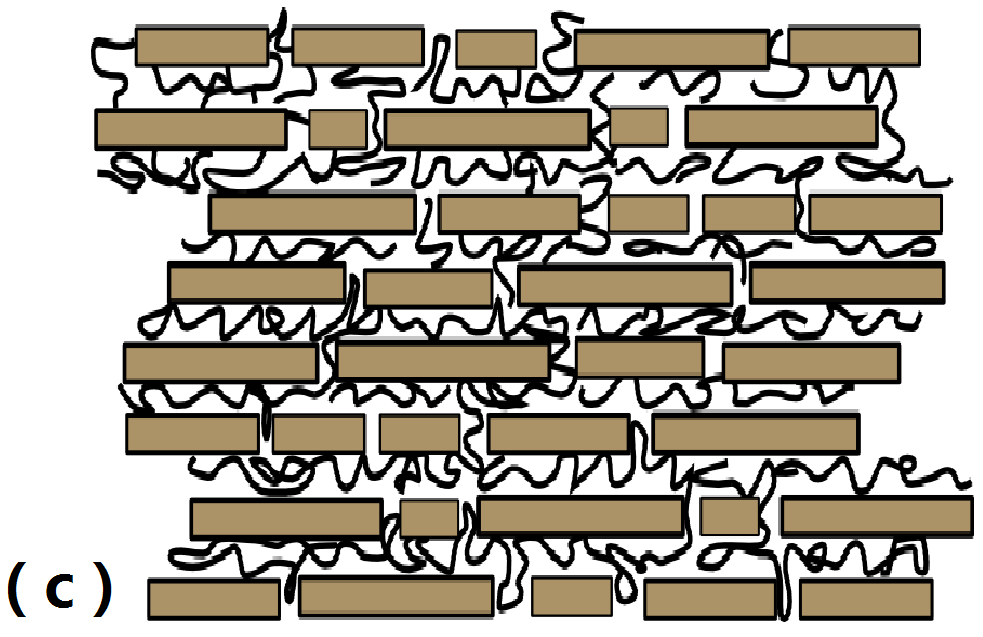}
       \caption{a) Picture of nacre inside a nautilus shell; b) Electron microscopy image of a fractured surface of nacre; c) Schematic of the microscopic structure of nacre; All images are from Wikipedia, https://en.wikipedia.org/wiki/Nacre.}
       \label{fig:nacre}
   \end{figure}
    
On the other hand, studies on statistical aspects of the 'brick-and-mortar' structures are rare, especially in regard to the type and tail of the probability distribution of strength, and no undisputed theoretical basis has been established so far. For general quasi-brittle structures their failure probabilities are usually formulated by the weakest-link model and its well known limiting Weibull distribution is widely used. When the representative volume elements (RVE) are not negligible compared to the size of the whole structure, the finite weakest-link model \cite{BazPan06, BazPan07, LeBaz09,LeBaz11,BazLe17} is sometimes considered.
    
Meanwhile, limitations of classical Weibull-type approach have long been realized \cite{BazLe17,Baz05,BazPla98,BazXi91}, and applicability of the weakest-link chain model has bee questioned, e.g., by Bertalan \cite{Ber14}. According to this model, the maximum load (which is, under load control, the failure load) occurs as soon as one representative volume of material (RVE) fails. This neglects the parallel couplings due to lateral interactions and the stress redistribution causing the release of stored energy due to crack growth. Additionally, Monte Carlo simulations show \cite{Ber14} the failure probability, $P_f$, to deviate from the Weibull distribution for small nominal stress
. Yet, in structural design, the lower tail of $P_f$ that is most important. 
    
Unlike the weakest link model, fiber bundle model takes account of the effect of stress redistribution by considering various load sharing rules among fibers. The fiber bundle model with the rule of equal load sharing was studied rigorously by Daniels \cite{Dan45}, who showed that its failure probability converges to the normal (or Gaussian) distribution as the number of fibers tends to infinity. 

Later, a chain-of-bundles model was formulated Harlow and Phoenix \cite{Har78I,Har78II} to obtain the failure probability of fibrous composites under uniaxial tension. They assumed that cross sections of the composite with certain finite spacing along the loading direction are statistically independent. So these cross sections are related to each other through series coupling in a chain and the individual cross sections are treated as bundles with some hypothetical local load-sharing rule. A significant deviation of $P_f$ from Weibull distribution is predicted. Due to overwhelming obstacles to large scale simulation at that time (1978), they analyzed only bundles with less than 10 fibers, and no verification through Monte Carlo simulations was given.
    
The new idea of this paper is to model the essence of failure behavior of the 'brick-and-mortar' structure of nacre under uniaxial tension by a brittle square fishnet pulled along one of the diagonals. Similar to the weakest link model, we obtain the failure probability of fishnet by calculating its counterpart---the survival probability, $1-P_f$. We consider not only the probability that every link would survive, but also the probabilities that of the fishnet remaining safe after some links have failed. These additional probabilities are then identified to be the cause of deviation of $P_f$ from the weakest-link model, producing Weibull distribution at the lower tail of $P_f$. 

    
The results of Monte Carlo simulations offer a concrete verification to the analytical derivation fishnet statistics theory and provide fruitful insights. In previous studies of electrical circuits,
Monte Carlo simulations have mostly followed the random fuse model (RFM) \cite{Ala06,Ber14}, in which the elastic structure is simplified as a lattice of resisters with random burnout thresholds. The RFM simulates the gradual failure of resister networks under increasing voltage, which is similar to the failure process of brittle elastic materials under uniaxial loading. Here, however, the emphasis in on the finite element method (FEM) with random element (or link) strength because it reflects better the mechanical failure of fishnet. To obtain a sufficient number of samples ($>10^5$) within a not too long run time, a simple finite element code is here developed, in Matlab. A unprecedented number,  $10^6$, is obtained within a few days of computation. With such a large number ($10^6$) of samples, the resulting histograms become visually identical to the theoretical cumulative probability density function (cdf) of failure probability $P_f$.
    

    \section{\large Statistical Modeling of Fishnet Structure}

    In nature, the mechanical responses of hydrated and dry nacres are quite different: Hydrated nacres like those in the shell of pearl oysters and abalones exhibit strong nonlinearity and a yielding plateau under tension, compression and shear \cite{WangSuo01}. However, dry nacres are mostly brittle and the ductility is vanishingly small under uniaxial tension as well as compression. In this paper, 
only a dry (and brittle) nacre is considered. Its behavior up to brittle failure may be treated as linearly elastic.

    \subsection{Equivalent Embedded Structure of Nacre: Fishnet}

   \begin{figure}[!h]
       \centering
       \includegraphics[width=0.8\textwidth]{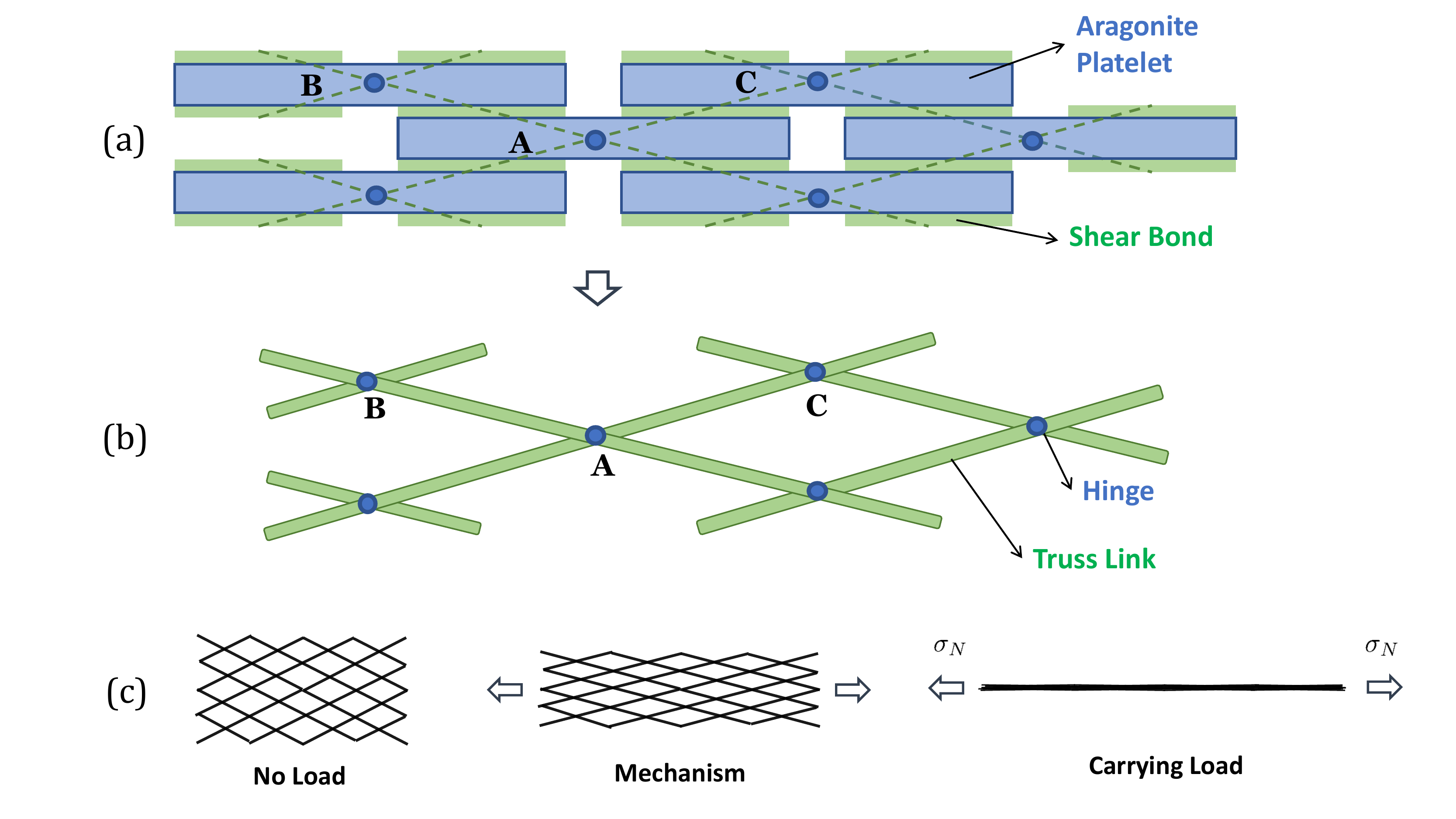}
       \caption{a) Schematic illustration of original nacre micro-structure; b) Equivalent fishnet structure; c) Deforming mechanism of fishnet}
       \label{fig:topology equivalence}
   \end{figure}
    To simplify and capture the essence of load transmission in the staggered imbricated structure of nacreous materials, we ignore the relatively weak tensile bonds providing the longitudinal connections of the adjacent lamellae (or platelets) and assume that most of the strength and stiffness is provided by the shear bonds between parallel lamellae (Fig.~\ref{fig:topology equivalence}(a)). Next, if we focus on the longitudinal strength of nacre under tensile load parallel to the lamellae, the bending stiffness of the thin polymer layers providing transverse connections of the lamellae does not play a significant role. So, for failure under longitudinal tension, we may imagine the lamellae to be replaced by their centroids, shown as the black dots in Fig. \ref{fig:topology equivalence} (a)), with the centroids, or nodes, connected by bars, or links (dashed lines in Fig. \ref{fig:topology equivalence} (a)), that transmit only axial forces. 

Thus we obtain a system of nodes and links shown in Fig. (\ref{fig:topology equivalence} (b), equivalent to a truss. Obviously, this truss looks like a fishnet. For example, in Fig.\ref{fig:topology equivalence}(a) the lamellae (or platelet) A,B and C are mapped to nodes A,B and C in Fig. \ref{fig:topology equivalence}(b) and the shear bonds AB and AC in Fig.\ref{fig:topology equivalence}(a) are replaced by truss elements AB and AC in Fig.\ref{fig:topology equivalence}(b), respectively. This is equivalent to pulling an elastic fishnet pulled along the longitudinal diagonals of the parallelogram cells. 

The truss is not statically determinate. Each of the parallelograms in the fishnet forms a mechanism which will, under longitudinal tension, shrink immediately at the start of loading into a set of coinciding lines (or fibers) with staggered connections among them (Fig.~\ref{fig:topology equivalence}(c)). Evidently, this is a simplification, which nevertheless retains the essence of transverse interactions between the adjacent rows of lamellae. 
For the purpose of illustration, the stress field in the fishnet will always be displayed ini the original (Lagragian) coordinates, i.e., before the collapse of fishnet into coinciding lines.  

The stress redistribution captured by the fishnet model plays a key role in the failure probability of fishnet. In this regard, the fishnet is different from each of the other two basic models for the statistics of failure, namely the weakest link chain model and the fiber bundle model.
    
    \subsection{Stress Redistribution Near a Crack in the Fishnet
    } 

The failure probability of a fishnet obviously depends on the stress redistributions after successive link failures. As an approximation for an analytical model, we calculate such redistributions deterministically, using a continuous approximation of the fishnet.
    
First we consider the failure of only one link. We assume the strength distributions of all links to be independent and identically distributed (i.i.d.) random variables (this implies that the autocorrelation length of the random strength field is assumed to be equal to the length of the link). The perimeter length of a set of geometrically similar fishnet structures (or domains) is proportional to the structure size, $D$, while the area of the fishnet structures is proportional to $D^2$. Thus the probability of the failed link to be located exactly on the boundary tends to zero as $D \to \infty$. Therefore, for large enough fishnets, the first link to fail lies, with probability almost 1, in the fishnet interior. So we study only the case of link failures in the fishnet interior.
    
\subsubsection{Stress Redistribution in Equivalent Fishnet Continuum}

  \begin{figure}[!h]
   \centering
   \includegraphics[width=1.05\textwidth]{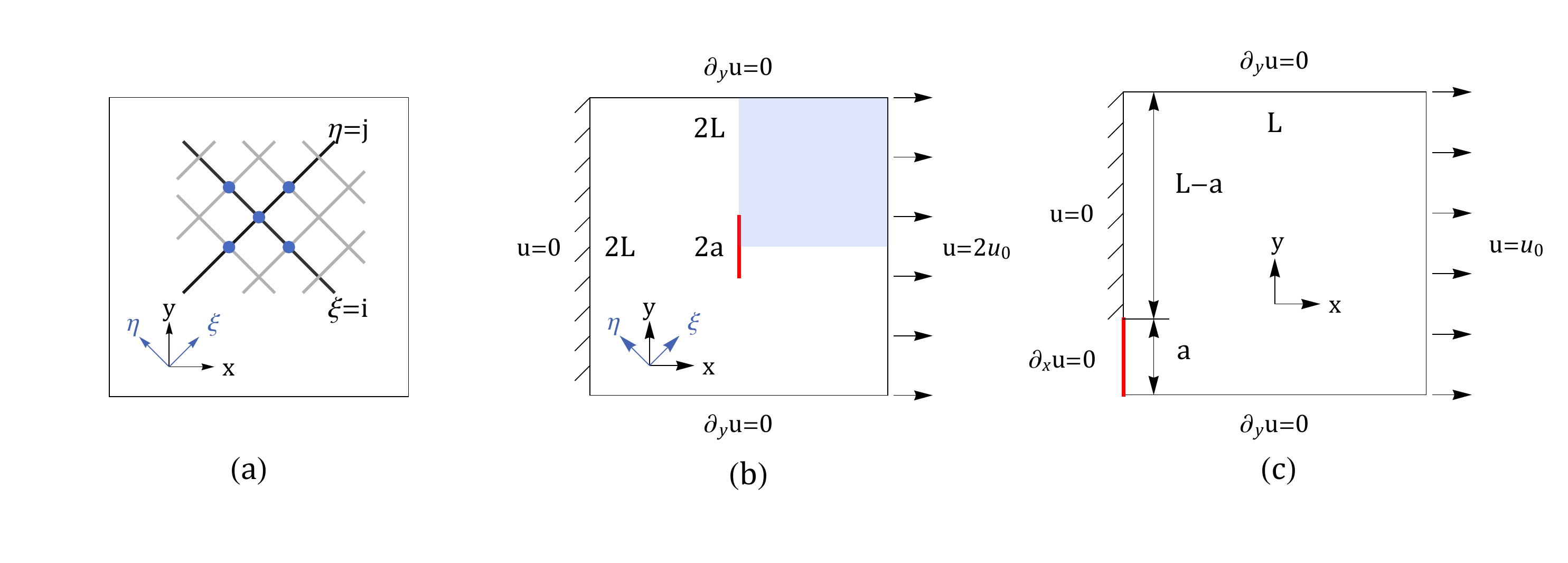}
   \caption{a) Schematic illustration of two orthogonal coordinate systems $x,y$ and $\xi,\eta$; b) Configuration of fishnet continuum with a pre-existing crack ; c) one quarter (shadowed region) of the fishnet continuum shown in (b)}
   \label{fig:discrete to continuum}
   \end{figure}
For simplicity, fishnets consisting of many links can be treated as a continuum. To formulate its governing equation, we consider a fishnet whose initial configuration consists of orthogonal lines of links (Fig.~\ref{fig:discrete to continuum}a), and introduce Lagrangian coordinates $(\xi,\eta)$ parallel to link lines. After the collapse of the fishnet into one line, all the four links connecting to one node $(i,j)$ become parallel, aligned in the $x$-direction. Since the initial fishnet state is kinematically indeterminate, a mechanism, the equilibrium condition of a node must be written for the collapsed state of fishnet. Equilibrium requires that the sum of the forces in $x$ direction from the four links acting on node $i,j$ would vanish. This leads to the difference equation:      
   \begin{equation}
        \label{finite difference}
        \frac{E A}{a}\left( \frac{u_{i,j+1}-u_{i,j}}{a} + \frac{u_{i,j-1}-u_{i,j}}{a} + \frac{u_{i+1,j}-u_{i,j}}{a} + \frac{u_{i-1,j}-u_{i,j}}{a} \right) = 0
    \end{equation}
where $u_{i,j}$ the displacement in the $x$ direction of any joint $(i,j)$, $a$ = length of each link, $E$ = Young's modulus and $A$ = cross section area of the link.

To obtain the continuum limit of $a \to 0$, consider Lagrangian coordinates $\xi, \eta$ in the initial directions of the links before fishnet collapse (Fig.~\ref{fig:discrete to continuum}a). Obviously, the continuum limit is the partial differential equation:
   \begin{equation}
        \label{Laplace xi}
        \frac{\partial^2 u}{\partial \xi^2} + \frac{\partial^2
    u}{\partial \eta^2} = 0 \mbox{~~~or~~~} \nabla_{(\xi,\eta)}^2u = 0
   \end{equation}
which is the Laplace equation. Since the Laplacian $\nabla^2$ is invariant at coordinate rotations, the differential equation of equilibrium in the $x$ direction in the initial $(x,y)$ coordinates before fishnet collapse reads:
    \begin{equation}
        \label{Laplace x}
        \frac{\partial^2 u}{\partial x^2} + \frac{\partial^2 
         u}{\partial y^2} = 0 \mbox{~~~or~~~} \nabla_{(x,y)}^2u = 0
    \end{equation}
where $x$ is the direction of loading of the fishnet and $y$ is the transverse direction with no load.
   
Note that if we defined the $(\xi,\eta)$ coordinate system in the deformed configuration (in which all the links become parallel and coincident with one line), the governing equation would become $\dd^2 u/\dd x^2=0$, which is independent of transverse direction $y$. Then, in the collapsed configuration, we would have to distinguish the coincident link lines by numbers rather than by $y$. But this description would be inconvenient for graphical representation.
 
The failure of one link or one group of links causes stress redistribution in the fishnet. Locally, many possibilities exist, which would complicate the calculation of failure probabilities. We assume that the redistributions can be characterized in the overall sense based on the continuum defined by Eq. (\ref{Laplace x}). The simplest case is a failure of a group of links approximated by a circular hole of radius $r_0$. Converting this partially differential equation to polar coordinates $(r, \tht)$, we have $(\pa^2 u / \pa r^2) + (1/r) (\pa u / \pa r) = 0$, and seeking an axisymmetric solution, one can readily find that the link strains $\pa u/\pa r$ as well as the stresses decay from the hole in proportion to $1/r$, according to the continuum approximation. 

In the case of a sharp crack of length $2a$ in the continuum, we may seek the displacement near the crack tip to be in the separated form $u = f(\tht) r^\la$ where $r$ is the distance from the crack tip. Upon substitution into Eq. (\ref{Laplace x}), it is then found that, like in linear elastic fracture mechanics (LEFM), the strains and stresses near the crack tip (i.e., for $r \to 0$) decay as $r^{-1/2}$ while far away from the crack (i.e., for $r \to infty$) they decay as $r^{-2}$. What is important is that the disturbance due to failed links is localized and decays to a negligible value within a certain finite distance. 
    Around the crack tip, the regions near the crack surface are shielded from the load an are unloaded to near zero. Later these regions will be seen to complicate the probabilistic analysis, compared to the the classical weakest-link and fiber-bundle models. This is the trade-off that we must accept to obtain a more accurate tail of failure probability of nacreous materials.
    
    \begin{figure}[!h]
       \centering
       \includegraphics[width=0.7\textwidth]{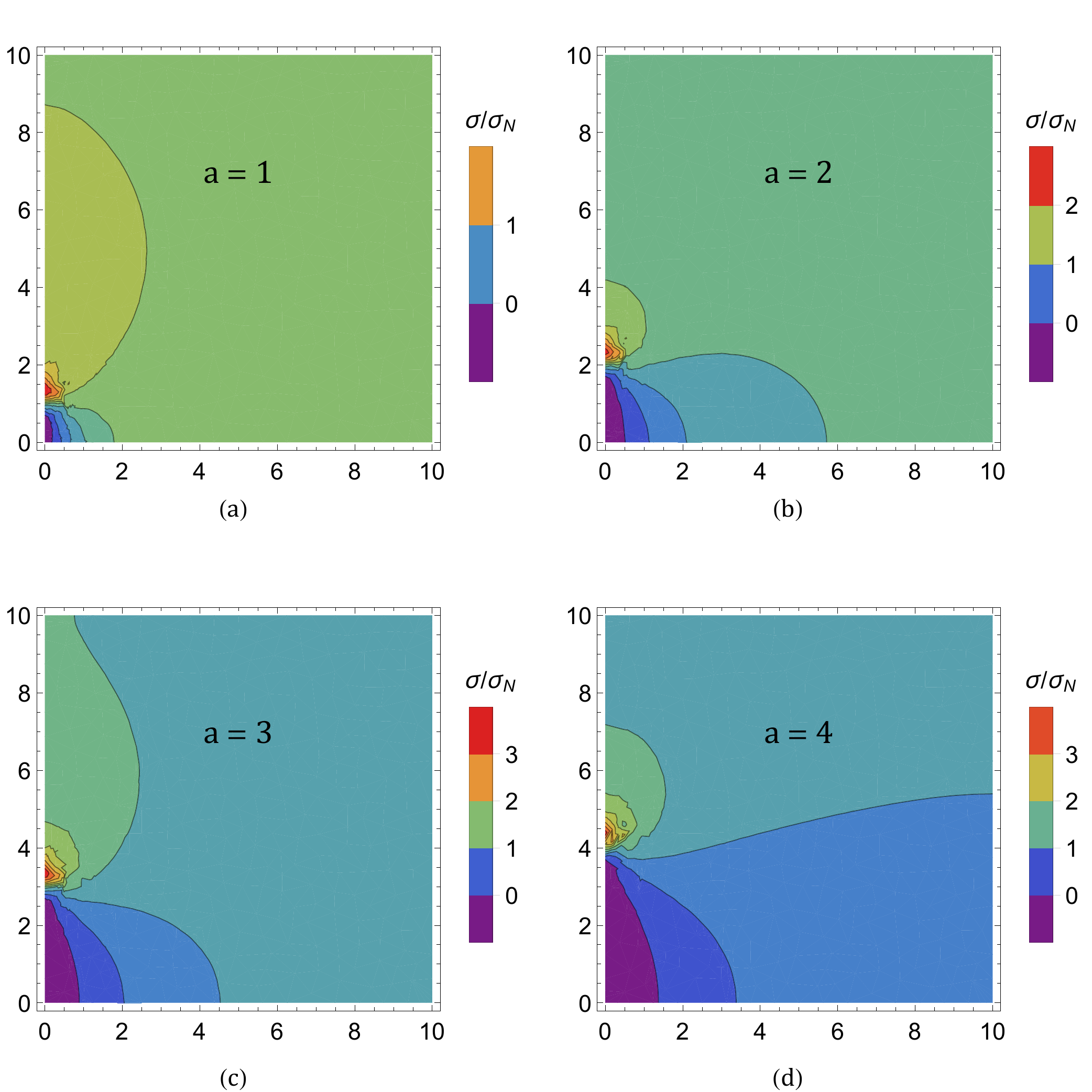}
       \caption{Profile of $\eta=\sigma/\sigma_N$ of fishnet continuum for different crack lengths: $a=$ 1, 2, 3 and 4. In this case, $L=10$.}
    \label{fig:continuum stress redistribution}
    \end{figure}   
    
\subsubsection{Stress redistribution in the     
fishnet}

    Knowing the asymptotic stress distribution near a crack (i.e., near one or more failed links) but, to calculate the failure probability accurately, it helps to calculate the exact discrete stress field. This is easily done by finite element method; see the results in Fig.~\ref{fig:discrete stress redistribution}.
    
    Fig.\ref{fig:discrete stress redistribution}a shows the stress field in a square fishnet with $64 \times 64$ links under uniform horizontal tension with a single failed link on the inside. The link properties are here assumed uniform and one link is assumed to be failed at the outset.  
Except close to this link, the stresses are almost uniform. To render the detail near the damage zone (inside the red square), Fig.\ref{fig:discrete stress redistribution}b presents a schematic showing the stresses in the damage zone only. We see that 8 close neighbors of the failed link endure stress redistributions with stresses greater than 10\% ($\eta_{max}=\sigma_{max}/ \sigma=1.6$ and $\eta_{min}=\sigma_{min} / \sigma = 0.64$. Four links with larger stress changes are highlighted by red and orange, and further four  with smaller stress changes by dark blue and blue).
The stress amplification zone is marked in light red, and the stress shielding (unloading) zone in light green.
    
As an example of how the stress ratio $\eta$ tends to 1 while moving away from the failed link, $\eta=\sigma/\sigma_N$ is plotted against the link number in Fig.\ref{fig:discrete stress redistribution}.c. In the figure, thresholds $|\eta-1|=5\%$ and $|\eta-1|=10\%$ are indicated by thick and dash lines respectively. Among 4095 links, $|\eta-1|>5\% $ occurs only on less than 20 links. The stress disturbance due to a failed link decays to a negligible value (less than 5\%) within the distance of 4 links. Later the exact values of $\eta$ near the crack tip are used to calculate the failure probability of fishnet.
    \begin{figure}[!h]
       \centering
       \includegraphics[width=0.9\textwidth]{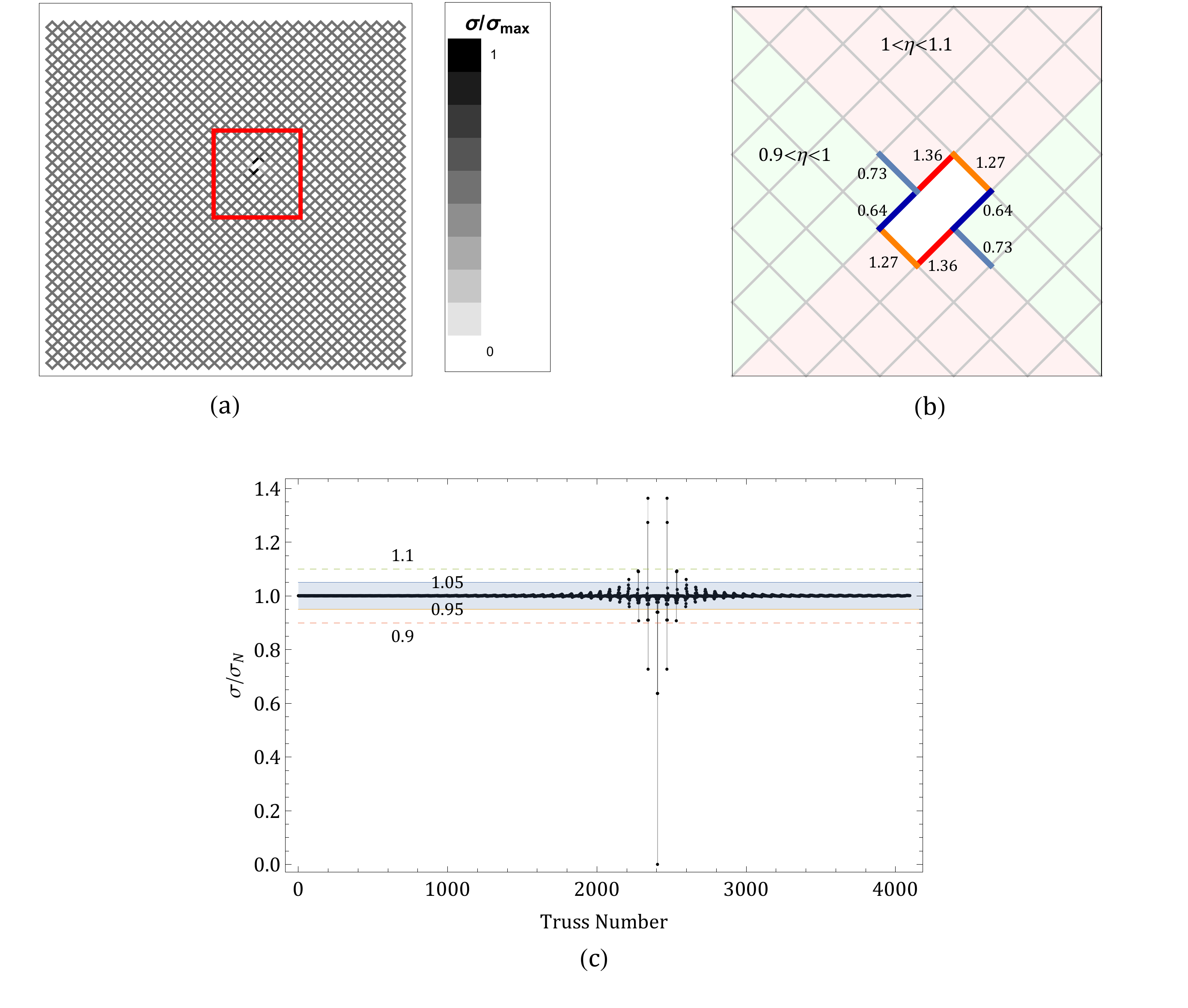}
       \caption{a) Normalized stress profile of fishnet with one failed link on the inside; b) Schematic showing the stress ratios ($\eta=\sigma/\sigma_N$) in the highlighted region of (a); c) Stress ratios of each link in (a) indexed by the link number in FE simulation.}
       \label{fig:discrete stress redistribution}
    \end{figure}


\subsection{Two Basic Models of Strength Statistics: Weakest-Link Chain and Fiber-Bundle}
    
Before embarking on fishnet statistics, we need to review the basics of this classical model for brittle failure. We consider a chain of $n$ identical links under tensile stress $\sig$ whose strengths represent independent identically distributed (i.i.d.) random variables. For the whole chain to survives, every single link must survive. So, according to the joint probability theorem,
    \begin{equation}
        \label{intro:chain}
        1-P_f(\sigma) = [1-P_1(\sigma)]^n   
    \end{equation} 
where $P_1$ and $P_f$ are the failure probabilities of a single link and of the whole chain, respectively. The limit of this equation for $n \to \infty$ may be written as
 \beq \label{e1}
  1-P_f(\sig) = \lim_{n \to 0} [1-x/n]^n = e^{-x},~~~x = n P_1(\sig) 
 \eeq
According to the stability postulate of extreme value statistics \cite{Fis28}, the type of the distribution for $\sig \to \infty$ must be invariant upon linear transformation of the independent variable $\sig$. One may check that this condition is verified if $P_1$ is a power function, $P_1(\sig) = \al \sig^{m_0}$, where $m_0$ and $\al$ are constants. According to the extreme value theorem \cite{Fis28}, the power function is the only possibility (among three) since the tensile strength $\sig$ cannot be negative. This leads to      
    \begin{equation}
        P_f(\sigma) = 1 - e^{-\alpha \sigma^{m_0}}
    \end{equation}
where $m_0$ is called the Weibull modulus and $\al$ is the scale parameter. 
    
For a chain with a finite number of links, only the left portion of $P_f(\sig)$ is Weibullian, and this portion shortens as $N$ decreases \cite{PangBazLe08}


Fiber composites as well as many other materials can often be treated, especially on a small scale, as bundle of fibers whose strengths are i.i.d. random variables. In contrast to a chain, the maximum load occurs only after the breakage of a certain number of fibers, which is generally larger than 1. After each fiber break, the load redistributes among the surviving fibers. Although the redistribution is different depending on the postpeak behavior of the fibers and stiffness of the loading platens, the total load distribution or on some empirical load-sharing rule. Asymptotically, the total load distribution of all fiber bundle models is Gaussian (or normal).


\subsection{Fishnet Statistics}

\subsubsection{Extension of Weakest Link Model}

    
    Consider a general 2D fishnet structure consisting of $N$ links whose failure probabilities $\sigma_i$ are i.i.d. random variables. 
In contract to a chain, more than one link may fail before the fishnet fails.
Therefore, we define the random variable $X(\sigma)$ 
= number of links that have failed after stress $\sig$ is imposed at the boundary. 
    
    Now consider a fishnet with $N=m \times n$ links, i.e., with $m$ rows and $n$ columns, pulled in the direction of rows. Since events $\{X(\sigma)=k\}$ are mutually exclusive, the whole probability space can be partitioned by these exclusive events into $N+1$ subsets: $\{X(\sigma)=k\}, k=0,1,\cdots, N$. Different from a chain or a fiber bundle, the failure of fishnet, however, cannot be characterized as either $\{X(\sigma)=1\}$, $\{X(\sigma)=m\}$ or $\{X(\sigma)=N\}$ because its failure also depends on the position of failed links, i.e. the whole structure fails when the $n_f$ failed links form section cutting through all the rows (vertically in Fig.~\ref{fig:nacre}(c)). So the condition $n_f \geq m$ is necessary for failure but not sufficient. The failed links may be scattered in the fishnet discontinuously.Based on these considerations, the probability of fishnet survival may be expressed as:
   \begin{align}   
   \label{general sum} \nn
   1-P_f(\sigma) =&   P_{S_0}(\sigma) + P_{S_1}(\sigma)
                 + P_{S_2}(\sigma) + \cdots + P_{S_{m-1}}(\sigma)\\  
                 &~~~ + \cPr( X(\sigma) \geq m~~ 
      \text{(the whole structure not yet failed )}
   \end{align}
where $P_f(\sigma) = (\sig_{max} \leq \sigma)$, $\sig_{max}$ = nominal strength of structure; and $P_{S_k}(\sigma) = \cPr(X(\sigma)=k),~k=0,1,2,...,N$. Since the event $\{X(\sigma)=0\}$ means that no link can fail under the load $\sigma$, we have
    \begin{equation}   \label{PS0}
     P_{S_0}(\sigma)=[1 - P_1(\sigma)]^n~~~\text{where}~~ 
     P_1(\sigma) = \cPr(\sigma_i \leq \sigma).
    \end{equation}
If we consider only the first term $P_{S_0}$ on the right-hand side of Eq.(\ref{general sum}), the model is equivalent to the classical weakest-link model,i.e., 
    \begin{equation}    \label{weakest link}
        1-P_f(\sigma) =  P_{S_0}(\sigma) = [1-P_1(\sigma)]^n
    \end{equation}
Physically, this means rearranging all the links into a chain with only 1 row and $N$ columns. Because the other terms in the sum of Eq.~(\ref{general sum}) are excluded, the weakest-link model gives a strict upper bound of failure probability of the fishnet. 

\subsubsection{
               Two-Term Fishnet Statistics}

To get a better approximation, let us now truncate the sum in Eq.~(\ref{general sum}) after the second term $P_{S_1}$. 
Let $\sig_N$ be the nominal stress representing the average stress in the cross section of the fishnet, and $\sigma_i$ be the stress at the $i^{th}$ link when the nominal stress is $\sigma_N=\sigma$. Also denote $\eta_j^{(1)} = \sigma_i /\sigma_N$, which is the ratio of stress change in link $i$ when some links failed. Based on our previous analysis of stress redistribution, we assume $\eta_j^{(1)}$ not to depend on the position of the failed link. For the two-term model, the survival probability is
    \begin{equation}
        \label{two term expansion}
        1-P_f(\sigma) =  P_{S_0}(\sigma) + P_{S_1}(\sigma),
    \end{equation}
where $P_{S_0}$ is given by Eq.(\ref{PS0}). and$P_{S_1}(\sigma)$ is the probability that, under load $\sigma_N=\sigma$, only one link has failed. This event can be further decomposed as the union of mutually exclusive events:
    \begin{equation}
        \{X(\sigma)=1\}=\cup_{i=1}^{N}\Omega_i,
    \end{equation}
where $\Omega_i$ is the event that the $i^{th}$ link is the only one that has failed under the loading $\sigma_N=\sigma$. Next, we define a family of events for every single link: 
    \begin{equation}
      \omega_i(\sigma) = \text{the $i^{th}$ link survives under 
                               stress} \sigma_i = \sigma
    \end{equation}
Then the event $\Omega_i$ that the $i^{th}$ link fails while the others survive is equivalent to
    \begin{equation}
        \Omega_i=\omega_i^c(\sigma) \cap \left\{ \bigcap_{j \neq i}
         \left[ \omega_j(\sigma) \cap \omega_j(\eta_j^{(1)}\sigma)
         \right] \right\}
    \end{equation}
where $\omega_i^c(\sigma)$ is the event that the $i^{th}$ link fails under the initial stress field $\sigma_i = \sigma$ and the expression $\bigcap_{j \neq i}\left[ \omega_j(\sigma) \cap \omega_j( \eta_j^{(1)} \sigma) \right]$ represents the event that the remaining links survive under both the initial and redistributed stresses. Hence the expression for $P_{S_{k}}$ is:
   \begin{align}
       \label{partition of PS1}
       P_{S_1}(\sigma) &=\cPr\{ X(\sigma)=1 \} =
   \cPr\{\cup_{i=1}^{N}\Omega_i\} = \sum_{i=1}^{N}  \cPr\{\Omega_i\} 
   \\  &=N\cdot\cPr\left\{ \omega_i^c(\sigma) \cap \left[ \bigcap_{j
     \neq i}\left( \omega_j(\sigma)\cap \omega_j(\eta_j^{(1)}\sigma)
      \right) \right] \right\}
   \end{align} 
Assume the strength of links $P_1(\sigma) = \cPr\{\omega_i (\sigma)\}$ are i.i.d. random variables. Then Eq.~(\ref{partition of PS1}) reduces to
   \begin{align}
   \label{PS1 if statement}
       P_{S_1}(\sigma) 
       &= N \cdot \cPr\{\omega_i^c(\sigma)\} \cdot \prod_{j\neq 
        i}\cPr\left\{ \omega_j(\sigma)\cap 
         \omega_j(\eta_j^{(1)}\sigma)  \right\}\\
       &= N P_1(\sigma)\cdot \prod_{j\neq i}\cPr\left\{ 
          \omega_j(\sigma)\cap \omega_j(\eta_j^{(1)}\sigma)  \right\}
   \end{align}
When $\eta_j^{(1)} \geq 1$, the event $\left\{ \omega_j(\sigma)\cap \omega_j(\eta_j^{(1)}\sigma) \right\}$ is equivalent to $\omega_j (\eta_j^{(1)}\sigma)$, and when $\eta_j^{(1)} < 1$, it is equivalent to $\omega_j(\sigma)$. Therefore, the expression for $P_{S_1}$ further reduces to
    \begin{equation}
    \label{exact expression of PS1}
       \boxed{P_{S_1}(\sigma)=NP_1(\sigma)\cdot \prod_{i=1}^{N-1}
        \left[ 1 - P_1(\lambda_i \sigma) \right],~~~\lambda_i=
           \begin{cases}
               \eta_i~,& \eta_i\geq 1\\
               1~~,& \eta_i<1
           \end{cases}}
    \end{equation}
This equation has been derived without using any knowledge of stress redistribution or any assumptions about load sharing rules. To get a simpler estimate of $P_{S_1}$, we can make two reasonable simplifications of stress redistribution:  
    \begin{enumerate}
        \item [1.] The stress redistributes only in the vicinity of a failed link. So we assume that links farther away are undisturbed, with all $\eta_i^{(1)} = 1$, and assume the stress to change only a few links, whose number is denoted as $\nu_1$ (not including the failed link);
        \item [2.] To keep $P_f$ either as an upper bound or an optimum prediction, we replace all the $\nu_1$ redistributed stresses $\eta_{i}^{(1)} \sigma$ by the maximum $\eta_{max}^{(1)}  \sigma$ (or by some weighted average $\eta_{a}^{(1)} \cdot \sigma$ among them).
    \end{enumerate}
Based on these two additional assumptions, $P_{S_1}$ is further simplified to
    \begin{equation}
    \label{simplified PS1}
       \boxed{P_{S_1}(\sigma) = N P_1(\sigma)[1 - P_1(\sigma)]^{N - \nu_1 - 1}[1 - P_1(\eta_a^{(1)}\sigma)]^{\nu_1}}
    \end{equation}
    
As a reasonable estimate (to be checked by Monte Carlo simulations), we choose $\nu_1$ such that all the $\eta_i^{(1)}$'s that are smaller than 1.1 would be replaced by 1, and then determine the value of $\eta_{a}^{(1)}$ by approximating Eq.~(\ref{exact expression of PS1}) by Eq.~(\ref{simplified PS1}), i.e.,
    \begin{equation}
        \prod_{i=1}^{N-1} \left[ 1 - P_1(\lambda_i \sigma)
         \right]\simeq [1-P_1(\sigma)]^{N - \nu_1 - 1}[1 - 
            P_1(\eta_a^{(1)} \sigma)]^{\nu_1}
    \end{equation}
    Based on calculation, for the fishnet we are studying right now and various strength distribution $P_1(\sigma)$ of links, $\nu_1=4\sim 8$ and $\eta_{a}^{(1)}=1.34 \sim 1.36$. From this, we can see that $\eta_{a}^{(1)}$ is very close to $\eta_{max}^{(1)}$, meaning that using the maximum redistributed stress can actually give us a pretty good prediction of $P_f$.
    
    To get the failure probability, we plug Eq.~(\ref{simplified PS1}) into Eq.~(\ref{two term expansion}):
   \begin{align}
       1-P_f(\sigma) &= [1-P_1(\sigma)]^N + 
       N P_1(\sigma)[1 - P_1(\sigma)]^{N - \nu_1 - 1}[1 - 
        P_1(\eta_a^{(1)} \sigma)]^{\nu_1}  \\
       & = [1 - P_1(\sigma)]^N \left\{ 1 + \frac{NP_1(\sigma)}{1 -
        P_1(\sigma)}  \left[\frac{1 - P_1(\eta_a^{(1)}\sigma)}{1 -
         P_1(\sigma)}\right]^{\nu_1}  \right\}
   \end{align}
Denoting 
    \begin{equation}
        P_{\Delta} 
        = \frac{1}{1-P_1(\sigma)}  \left[\frac{1 - 
           P_1(\eta_a^{(1)}\sigma)}{1 - P_1(\sigma)}\right]^{\nu_1}
    \end{equation}
we finally get
    \begin{equation}
        \label{extension}
       \boxed{1-P_f(\sigma) = [1-P_1(\sigma)]^N \cdot \left\{ 1 + N
        P_1(\sigma) P_{\Delta}(\sigma,\eta_a^{(1)},\nu_1) \right\}}
    \end{equation}
    From Eq.~(\ref{extension}), we can see that the main difference between weakest link model with the two-term fishnet model is the extra term:
    \begin{equation}
        N P_1(\sigma) \cdot P_{\Delta}
    \end{equation}

In the expression for $P_{\Delta}$, $[(1-P_1(\eta_a^{(1)} \sigma))/(1-P_1(\sigma))]^{\nu_1}$ is the conditional probability of survival for the links in the region of stress redistribution with load load $\eta_a^{(1)} \sigma$, given that they have been carrying load $\sigma$ without failing. In other words,  
    \begin{equation}
        1-\left[\frac{1-P_1(\eta_a^{(1)}
         \sigma)}{1-P_1(\sigma)}\right]^{\nu_1}
    \end{equation}
is a cumulative distribution function (cdf) of structural failure probability. Therefore, its complement has the following properties:
   \begin{equation*}
       \lim_{\sigma \rightarrow 0}\left[\frac{1-P_1(\eta_a^{(1)} 
                   \sigma)}{1-P_1(\sigma)}\right]^{\nu_1} = 1
       ~~~\text{and}~~~
       \lim_{\sigma \rightarrow \infty}\left[\frac{1-P_1(\eta_a^{(1)} \sigma)}{1-P_1(\sigma)}\right]^{\nu_1} = 0
   \end{equation*}
    As for the other term in $P_{\Delta}$, i.e. $1/(1-P_1(\sigma))$, it tends to 1 as $\sigma$ tends to zero and blows up as $\sigma$ approaches $\infty$. Therefore as $\sigma$ tends to 0, the term $P_{\Delta}$ tends to 1. 
Numerical study of $P_{\Delta}$ shows that for most cases with $P_1$ being Gaussian, Weibull or other typical grafted distributions, $P_{\Delta}$ always tends to 0 as $\sigma$ approaches $\infty$ as long as $\eta_a^{(1)}>\eta_0$, where the constant $\eta_0$ depends on $P_1$ and is greater but very close to 1 (e.g $\eta_0=1.02$).
    
For structures such as bridges or aircraft, it is generally desired that the structural failure probability be less than $10^{-6}$, and in that case we obviously need to pay attention only to the remote left tail of $P_f(\sigma)$, i.e., for $\sig \to 1$. In that case we have
    \begin{equation}
        \label{two term simplification}
        1-P_f(\sigma) \simeq [1-P_1(\sigma)]^N 
        \left[ 1 + N P_1(\sigma) \right]
    \end{equation}
To elucidate the difference from the classical weakest link model, it helps to transform the foregoing equation (Eq.~(\ref{two term simplification})) to the Weibull scale, which has the coordinates:
    \begin{equation}
        X^*=\ln \sigma,~Y^*=\ln[-\ln(1-P_f)]
    \end{equation} 
For comparison, we plot in this scale also the weakest link model. For $\sig \to 0$, we have $P_1 \to 0$, and so we think of using the approximation $\ln(1+x) \simeq x$ for small $x$. However, with these approximations, the logarithm of Eq. (\ref{two term simplification}) gives 0. 
So we must use the second order approximation $\ln(1+x) \simeq x - x^2/2$. This leads to:
   \begin{align}
       \label{Taylor expansion}
       \ln[1 - P_f(\sigma)] 
       &\simeq N\ln[1 - P_1(\sigma)] + \ln[1 + N P_1(\sigma)]\\
       &\simeq N \left( - P_1 - \frac{P_1^2}{2} \right) + NP_1
        - \frac{(N P_1)^2}{2}\\
       & = - N(N+1)\frac{P_1^2}{2}
   \end{align}

The previous derivation shows that by adding the term $P_{S_1}$, the dominant term of $\ln[1-P_f]$ changed from $P_1(\sigma)$ to $[P_1(\sigma)]^2$, which will be immediately seen to have a huge effect on the tail of $P_f$. To obtain the Weibull scale plot, we multiply both sides by $-1$ and take again the natural logarithm; this yields:
    \begin{equation}
        \label{fishnet2}
        Y^* = \ln[-\ln[1-P_f(\sigma)]] = \ln 
         \frac{N(N+1)}{2} 
          + 2\ln P_1(\sigma)
    \end{equation}
For comparison, the Weibull plot for the weakest-link model is
    \begin{equation}
        \label{Weibull}
        Y^*=\ln[-\ln[1-P_f(\sigma)] ] = \ln N + \ln P_1
    \end{equation}
Since the independent variable of Weibull plot is $X^*=\ln\sigma$, and $\ln P_1 \simeq \ln(\sigma^{m_0}) = m_0 \ln \sigma $, the slope of the Weibull plot for the Weibull distribution is the Weibull modulus $m_0$. Comparing the expressions of Eq.~(\ref{fishnet2}) and Eq.~(\ref{Weibull}), one can find that the slope of the far left tail of Weibull plot in the two-term fishnet model is changed from $m_0$ into $2m_0$, which is twice as steep as in the weakest link model.
    
    \begin{figure}[!h]
        \centering
       \includegraphics[width=0.95\textwidth]{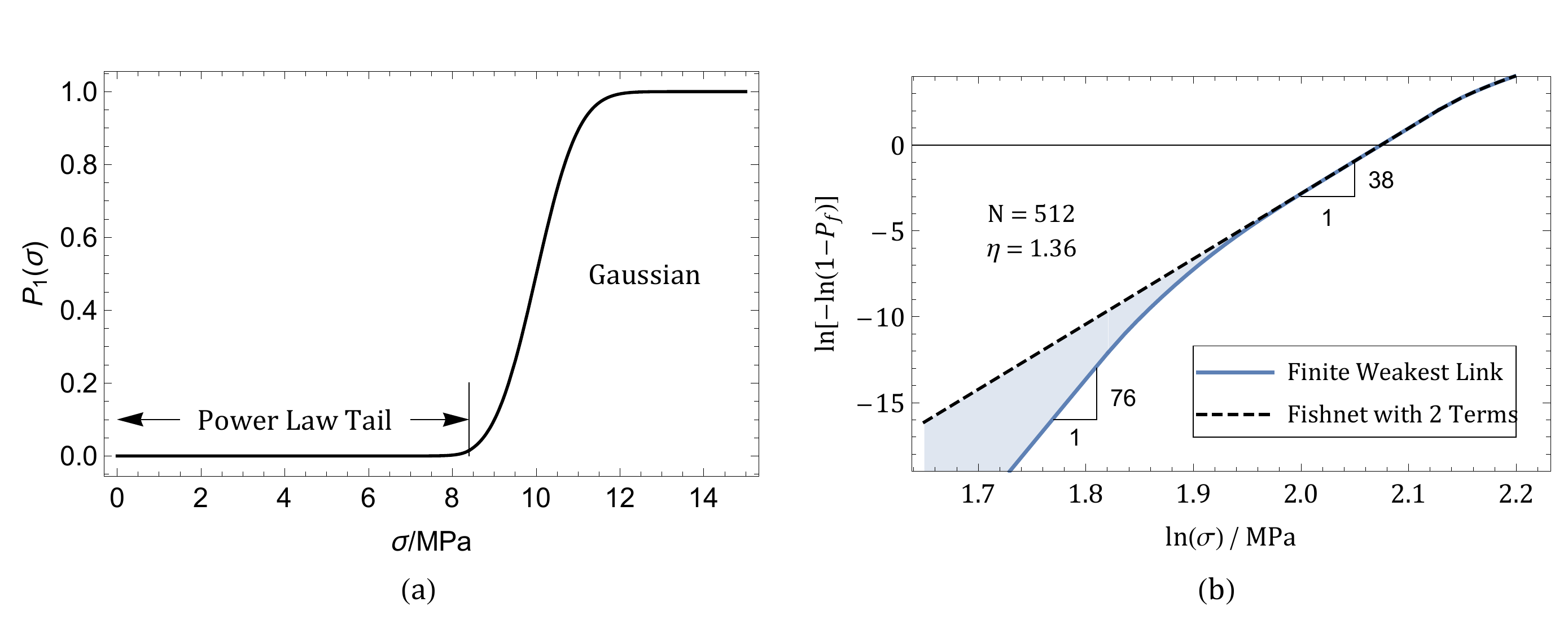}
        \caption{a) Cumulative distribution function (cdf) of failure for a single truss element; b) Comparison between finite weakest link model and fishnet model with 2 terms in expansion (Weibull scale).}
    \label{fig:analytical comparison}
    \end{figure}    
    
On the other hand, in the region where $P_{\Delta}$ is close to 0, i.e., for large enough $\sigma$, Eq.~(\ref{extension}) reduces to the weakest link model. Thus the addition of the term $P_{S_1}$ affects only the far left tail of the cdf of structural strength and has no effect on the right tail. Physically, this is because when $\sigma_N$ is very small (far below the mean), the links surrounding the failed link are generally strong enough to bear the redistributed load without breaking. But the weakest-link model characterizes this case as a failure of the whole structure, and this is why it overshoots $P_f(\sigma)$ when the load is small. However, when $\sigma_N$ is large (far above the mean), the survival probability of the links  surrounding the failed one is very small, regardless of the failed link. So, on the far right tail, the two-term fishnet model is essentially the same as the weakest-link model.

Fig.~\ref{fig:analytical comparison}(b) shows the difference between the finite weakest-link model and the two-term fishnet model for the case in which $N=512$, $\eta_a^{(1)}=1.36$, $\nu_1=6$, and the cdf $P_1(\sigma)$ of structure strength is given in Fig.~\ref{fig:analytical comparison}(a). Note that $P_1(\sigma)$ is the Gaussian (or normal) distribution with a power law tail grafted at cumulative probability $P_1(\sigma)=0.015$. From Fig.~\ref{fig:analytical comparison}(b), one should note that the two-term fishnet model shows, in the Weibull plot, a smooth transition of slope from $m_0=38$ to $2m_0=76$. Note that, since the two-term fishnet statistics does not consider all the terms in Eq.~(\ref{general sum}), it is still an upper bound for the true $P_f(\sigma)$. 

The difference between the weakest-link model and the two-term fishnet model is illustrated by the diagram in Fig.\ref{fig:analytical comparison}(b), in which $\sigma = 6.05$ MPa. 
The probability calculated from the two-term fishnet model is $1.19 \cdot 10^{-6}$, whereas the value calculated from the weakest-link model is $2.95\times10^{-5}$. This is 24.8 times higher than the value obtained by the two-term fishnet model. This unexpected observation shows that when $\sigma$ is small, the weakest-link model gives a much too high prediction of failure probability, and the two-term fishnet  model yields much higher estimates of failure loads, provided that the fishnet action indeed takes place.
    
\subsubsection{Location of the Slope Transition for the 2-Term Fishnet Statistics}

It is not enough to know that the left terminal slope in the Weibull plot is doubled. One needs to know also at which $\sig$ value the slope transition is centered. According to Eq.~\ref{general sum}, this value is controlled by two parameters, $\eta_a^{(1)}$ and $\nu_1$. Here $\eta_a^{(1)}$ is the ratio of the average redistributed stress to the initial uniform stress, and $\nu_1$ is the number of links in the redistribution zone. To study the effects of these parameters, we fix one of them plot $P_{\Delta}$ for various values of the other. We consider $N = 512$ and $\nu_1 = 6$. The calculated plots of $P_{\Delta}$ and $P_f$ for $\eta_a^{(1)}=$ 1.1, 1.3 and 1.6 are given in Fig.~\ref{fig:Various Eta}.

    \begin{figure}[!h]
        \centering
       \includegraphics[width=0.95\textwidth]{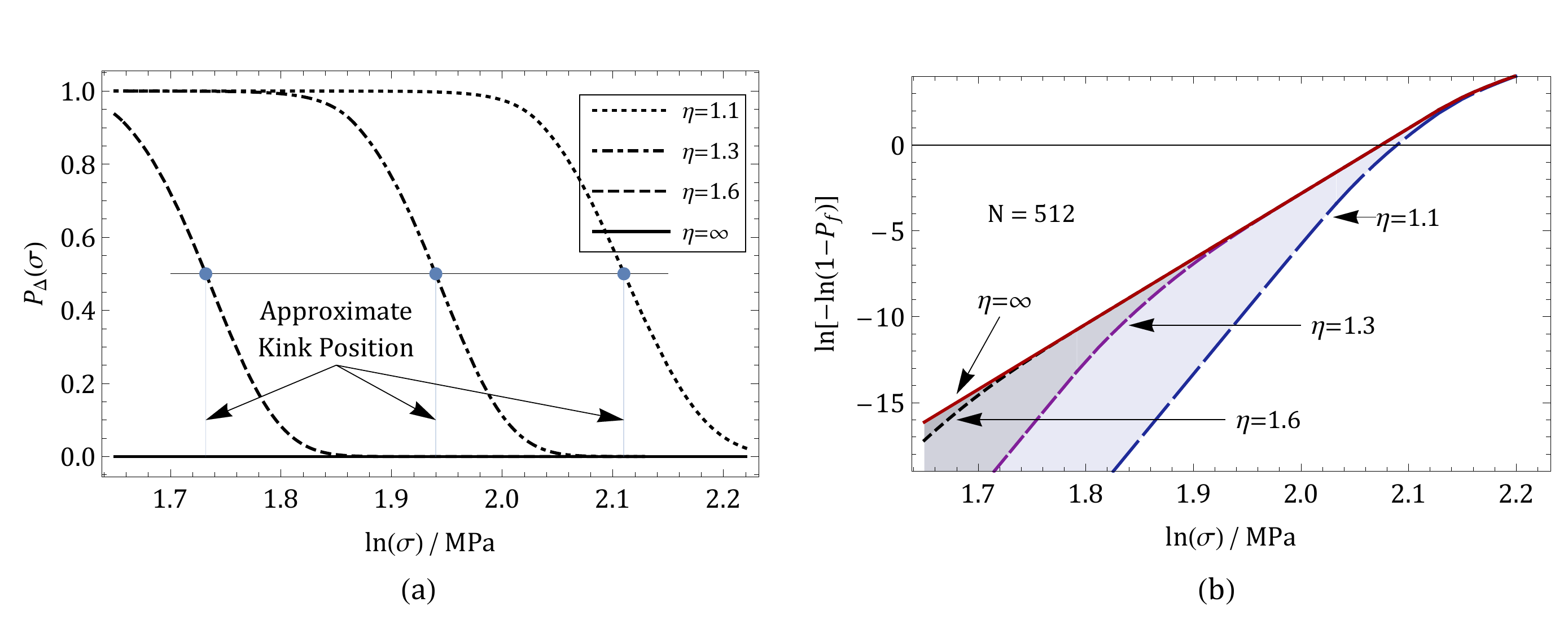}
        \caption{a) Approximate position of transition for $\eta_a^{(1)}=$ 1.1, 1.3 and 1.6; b) Cumulative distribution function (cdf) of $P_f(\sigma)$ in Weibull scale with different 5$\eta_a^{(1)}$.}
        \label{fig:Various Eta}
    \end{figure}
    \begin{figure}[!h]
        \centering
       \includegraphics[width=0.95\textwidth]{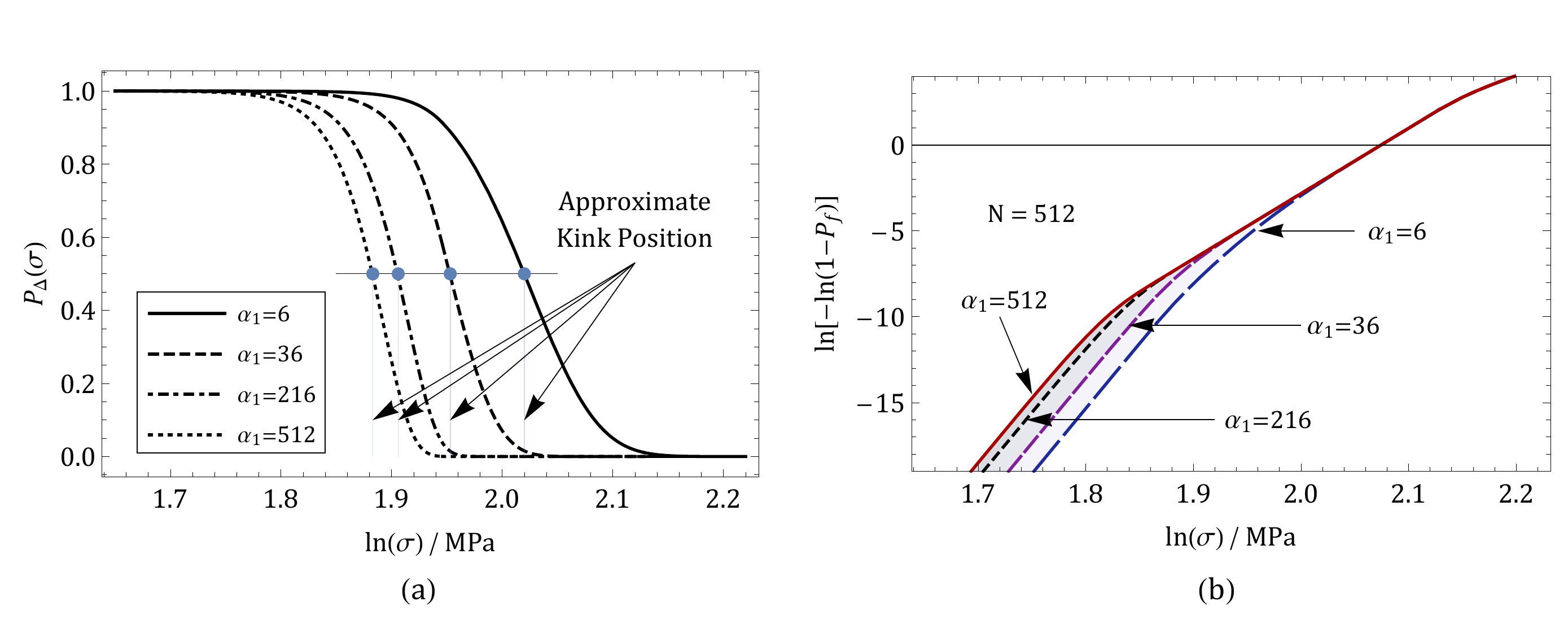}
        \caption{a) Approximate position of transition for $\nu_1=$ 6, 36, 216 and 512; b) Cumulative distribution function (cdf) of $P_f(\sigma)$ in Weibull scale with different $\nu_1$.}
        \label{fig:Various alpha1}
    \end{figure}

As the value of $P_{\Delta}$ varies from 0 to 1, the slope of cdf in Weibull scale gradually transits from $m_0$ to $2m_0$. Thus the center of the transition is the value $\sig_T$ of $\sigma$ at which $P_{\Delta}(\sigma_T)=0.5$. So we set $P_{\Delta}=0.5$ for various values of $\eta_a^{(1)}$, and compare the $\sigma_T$ values. Shown in Fig.~\ref{fig:Various Eta}(a), they are seen match well with the kink locations in Fig.~\ref{fig:Various Eta}(b). Also note that the kink location shifts dramatically to the left when $\eta_a^{(1)}$ is increased from 1.1 to 1.6. Finally, as $\eta_a^{(1)} \to \infty$, the kink position approaches $\sigma_T=0$, which means that the fishnet degenerates into the weakest-link model---as long as 1 link fails, the stress in its neighborhood becomes infinitely large, then the structure must fail right after a single link fails.
    
Next, we fix $\eta_a^{(1)} = 1.2$ and let $\nu_1$ = 6, 36, 216 and 512. Note that, since $N = 512$, the largest $\nu_1$ value cannot exceed $512$. Similarly, as $\nu_1$ increases, the kink location shifts to the left, but the shift is much smaller than the change of $\eta_a^{(1)}$. Thus we conclude that the stress ratio $\eta_a^{(1)}$ has a big effect on the location of slope change from $m_0$ to $2m_0$, while $\nu_1$ has a much smaller effect.
    
 
    \subsubsection{Three-Term Fishnet Statistics} 

After a link fails, the next one to fail will be either a neighboring link or a link located farther away. Since these two events are mutually exclusive, we may write:
    \begin{equation}
    P_{S_2} = P_{S_{21}} + P_{S_{22}},
    \end{equation}
where $P_{S_{21}}$ (or $P_{S_{22}}$) is the probability of $\{X = 2\}$ when the failed link is close to (or far away from) the previously failed link. Furthermore, the probability density function (pdf) characterizing $P_1(\sigma)$ is denoted as $\psi(\sigma) = \dd P_1(\sigma) /\dd \sigma$. If the second link failure after the original link failure happens in the neighborhood of the first one, we let $\nu_2$ be the number of links that will endure the redistributed stresses after 2 links (after the original one) fail and let the new stress ratio be $\eta^{(2)}~(note that \eta^{(2)}>\eta_a^{(1)}>0)$. In both foregoing cases, since there are only 2 failed links (after the original one), we still assume that the stress in the links far away from the failed ones remains undisturbed and equals the remotely applied stress $\sigma$ (this ceases to be valid if number of failed inks keeps increasing further). Thus we may write:
    \bea
    \label{formulation of PS21}
    P_{S_{21}} &=& {N \choose 1}{\nu_1 \choose 1}\int_{0}^{\sigma}\int_{x_1}^{\eta_b^{(1)} \sigma}\psi(x_1)\psi(x_2)dx_2dx_1\
    [1-P_1(\sigma)]^{N-\nu_2-2}\ [1-P_1(\eta^{(2)} \sigma)]^{\nu_2}
    \\ 
    \label{formulation of PS22}
    P_{S_{22}} &=& {N \choose 1}{N-\nu_1-1 \choose 1} \int_{0}^{\sigma}\int_{x_1}^{ \sigma}\psi(x_1)\psi(x_2)dx_2dx_1\
    [1-P_1(\sigma)]^{N-2\nu_1-2}\ [1-P_1(\eta^{(2)}\sigma)]^{2\nu_1}~~    
    \eea
where $x_1$ is .... and  
$\eta_b^{(1)}$ is a parameter calculated from $\eta_i^{(1)}$. This parameter is used by another approximate relation which is similar but different from that for $\eta_a^{(1)}$:
    \begin{equation}
    \label{new approximation}
        \nu_1\int_{0}^{\sigma}\int_{x_1}^{\eta_b^{(1)} \sigma}\psi(x_1)\psi(x_2)dx_2dx_1 \simeq \sum_{j=1}^{\nu_1}\int_{0}^{\sigma}\int_{x_1}^{\eta_j^{(1)} \sigma}\psi(x_1)\psi(x_2)dx_2dx_1
    \end{equation}
 
    \begin{figure}[!h]
        \centering      
        \includegraphics[width=0.85\textwidth]{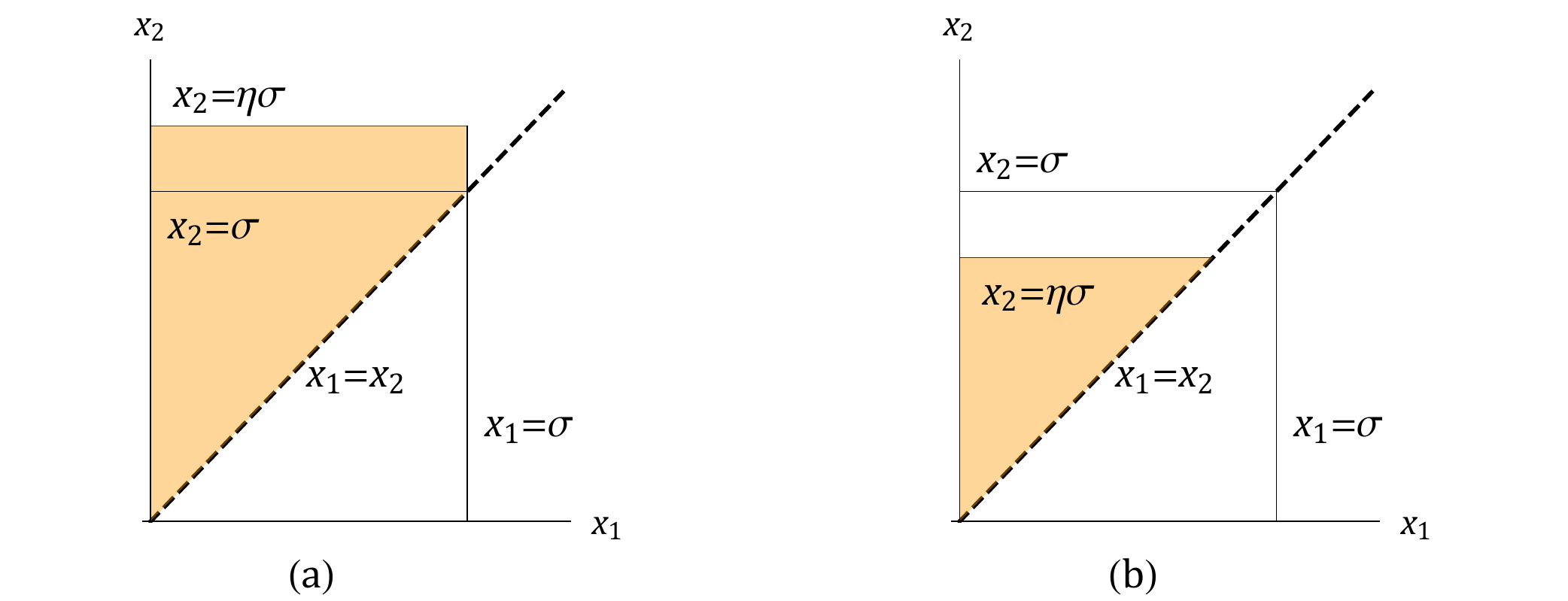}
        \caption{a) Schematic showing the domain of integration when $\eta \geq 1$; b) domain of integration when $\eta<1$.}
        \label{fig:integral schematic}
    \end{figure}

    Further note that
    \begin{equation}
    \int_{0}^{\sigma}\int_{x_1}^{ \sigma}\psi(x_1)\psi(x_2)dx_2dx_1=\frac{1}{2}[P_1(\sigma)]^2
    \end{equation}
    and then, according to Fig.~\ref{fig:integral schematic}(a),
   \begin{align}
   \int_{0}^{\sigma}\int_{x_1}^{\eta_b^{(1)} \sigma}\psi(x_1)\psi(x_2)dx_2dx_1 
   &= \int_{0}^{\sigma}\int_{x_1}^{ \sigma}\psi(x_1)\psi(x_2)dx_2dx_1
   +\int_{0}^{\sigma}\int_{x_1}^{\eta_b^{(1)} \sigma}\psi(x_1)\psi(x_2)dx_2dx_1\\
   &=\frac{1}{2}[P_1(\sigma)]^2+P_1(\sigma)[P_1(\eta_b^{(1)} \sigma)-P_1(\sigma)]\\
   &=P_1(\sigma)P_1(\eta_b^{(1)} \sigma)-\frac{1}{2}[P_1(\sigma)]^2
   \end{align}
Finally, we obtain
   \begin{align}
   \label{3-term fishnet}
   P_{S_{21}} &= N \nu_1 \left\{ P_1(\sigma)P_1(\eta_b^{(1)} \sigma)-\frac{1}{2}[P_1(\sigma)]^2 \right\}\cdot[1-P_1(\sigma)]^{N-\nu_2-2} \cdot [1-P_1(\eta^{(2)} \sigma)]^{\nu_2}\\
   P_{S_{22}} &= \frac{1}{2}N (N-\nu_1-1) [P_1(\sigma)]^2\cdot
   [1-P_1(\sigma)]^{N-2\nu_1-2} \cdot [1-P_1(\eta^{(2)} \sigma)]^{2\nu_1}
   \end{align}
So, considering three terms in Eq.~\ref{general sum}, we find the failure probability of fishnet to be
   \begin{align}
   1-P_f(\sigma) 
   &\simeq [1-P_1(\sigma)]^N \cdot \left\{ 1 + P_{S_1}/[1-P_1(\sigma)]^N + P_{S_2}/[1-P_1(\sigma)]^N \right\}\\
   &=[1-P_1(\sigma)]^N \cdot \left\{ 1 + NP_1(\sigma)P_{\Delta}
   + N\nu_2\left[P_1(\sigma)P_1(\eta_b^{(1)} \sigma)-\frac{1}{2}[P_1(\sigma)]^2\right] P_{\Delta_{21}}\right.\\
   & + \left. \frac{1}{2}N(N-\nu_1-1)[P_1(\sigma)]^2 P_{\Delta_{22}} \right\},
   \end{align}
where
    \begin{equation}
    \label{estimation of PS2}
    P_{\Delta} =  \frac{[1-P_1(\eta_a^{(1)} \sigma)]^{\nu_1}}{[1-P_1(\sigma)]^{\nu_1+1}},~~~
    P_{\Delta_{21}} = \frac{[1-P_1(\eta^{(2)} \sigma)]^{\nu_2}}{[1-P_1(\sigma)]^{\nu_2+2}},~~~
    P_{\Delta_{22}} = \frac{[1-P_1(\eta_a^{(1)} \sigma)]^{2\nu_1}}{[1-P_1(\sigma)]^{2\nu_1+2}}
    \end{equation}
Similar to $P_{\Delta}$, $\lim_{\sig \to 0} P_{\Delta_{21}} = 1$ and $\lim_{\sig \to 0} P_{\Delta_{22}} = 1$. 
    
Note that the foregoing expression for $P_{S_2}$is is not exact expression only an approximate estimate because we simplified, in two steps, the stress redistribution as a uniform scaling of stress by factors  $\eta_(b)^{(1)}$ and $\eta^{(2)}$. They are both greater than 1. In reality, however, some regions near the group of failed links get unloaded. These are regions in which the stress is smaller than the remotely applied stress $\sigma$. The failure probability of these regions is the integral of joint density over the domain shown in Fig.~\ref{fig:integral schematic}(b). This integral makes a more accurate solution very complicated, since stronger links could fail prior to the weaker ones if unloading happens in the weak links. Nevertheless, our estimation can still be quite accurate if $\eta_b^{(1)}$ is chosen properly, based on Eq.~(\ref{new approximation}).
    
    Since the probability $P_{S_2}$ consists of two terms, namely $P_{S_{21}}$ and $P_{S_{22}}$, another interesting question arises: which one of these two terms is larger?  Eq.~(\ref{formulation of PS21}) and Eq.~(\ref{formulation of PS22}) tell us that $P_{S_{21}}$ is proportional to $N[P_1(\sigma)]^2$, while $P_{S_{22}}$ is proportional to $N(N-\nu_1-1)[P_1(\sigma)]^2$, which approximately equals $N^2 [P_1(\sigma)]^2$. Therefore, $P_{S_{22}}$ is much larger than $P_{S_{21}}$. This means that the second link failure will most likely appear far away from the first one if the left tail of $P_1(\sigma)$ is relatively heavy but such that the effect of $[P_1(\sigma)]^2$ is still recognizable. This tendency to scattered damage will be discussed later.
    
\subsubsection{Ramifications to Higher-Order Terms in Failure Probability}

Intuitively, $P_{S_k}$ is proportional to $[P_1(\sigma)]^k$ when $\sigma$ is close to 0, and so Eq.~(\ref{general sum})is similar to a Taylor series expansion of some real function in terms of $P_1(\sigma)$. The higher the order, the smaller the contribution to the total sum. If the tail of $P_1(\sigma)$ is very light, $[P_1(\sigma)]^k$ for $k \geq 3$ are usually too small to be seen on the cdf directly. So, in order to see the effect of the higher order terms, we need, once again, to study them in the Weibull scale. This scale exponentially magnifies tiny differences in the tail of a distribution.
    
What will happen if we include more and more terms on the right-hand side of Eq.\ref{general sum}? Will they cause further slope increases in the Weibull plot? The answer is affirmative: The higher-order terms of survival probability would lead to further slope increase of cdf in the  Weibull scale. From the derivation of Eq.~(\ref{Taylor expansion}), we know that the very reason for slope increase by the  factor of 2 is that the term $P_1(\sigma)$ canceled out by Taylor expansion, allowing $[P_1(\sigma)]^2$ dominates. So, if there is another slope increase from $2m_0$ to $3m_0$, the square terms $[P_1(\sigma)]^2$ in Eq.~(\ref{Taylor expansion}) must vanish after adding more terms of $P_{S_k}$ in Eq.(\ref{general sum}). 
    
    In order to show that the slope of cdf could further increase due to the presence of more terms in the expansion of survival probability we must use the exact law of stress redistribution, otherwise it could lead to false conclusions. However, the exact law of stress redistribution is almost impossible to obtain analytically. Therefore, instead of working on a general fishnet, it is much easier if we consider a special form of it i.e. fiber bundle: by changing the aspect ratio $m:n$ gradually to $N:1$, a fishnet with a fixed number of links reduces to a fiber bundle. Then we can use the law of stress redistribution of a fiber bundle in derivation and study the slope change of cdf and the case for general fishnet is studied numerically by Monte Carlo simulations.
    
    Now, consider a fiber bundle consisting of $N$ fibers and we truncate Eq.(\ref{general sum}) at $P_{S_2}$. Consequently $P_{S_0}$, $P_{S_1}$ and $P_{S_2}$ can be expressed as:
    Similarly,
    \begin{equation}
    \int_{0}^{\sigma}\int_{0}^{N\sigma/(N-1)}\psi(x_1)\psi(x_2)dx_2dx_1=P_1(\sigma)P_1[N \sigma/(N-1)]-\frac{1}{2}[P_1(\sigma)]^2
    \end{equation}
    Therefore,
    \begin{equation}
        P_{S_2}=N(N-1)\left\{P_1(\sigma)P_1[N \sigma/(N-1)]-\frac{1}{2}[P_1(\sigma)]^2\right\}\left[ 1-P_1\left(\frac{N \sigma}{N-2}\right) \right]^{N-2}
    \end{equation}
    The probability of survival of the whole fishnet can then be expressed as:
    \begin{equation}
        1-P_f(\sigma) \simeq P_{S_0} + P_{S_1} + P_{S_2}
    \end{equation}
    Then we factor out $[1-P_1(\sigma)]^N$ and let $\sigma$ tends to 0:
    \begin{equation}
        1-P_f(\sigma) \simeq [1-P_1(\sigma)]^N \cdot \left\{ 1 + NP_1(\sigma) + N(N-1)P_1(\sigma)P_1[N \sigma/(N-1)]-\frac{N(N-1)}{2}[P_1(\sigma)]^2 \right\}
    \end{equation}
    By the same procedure, we take natural log on both sides and then use Taylor expansion of $\log(1+x)$ and collect all the terms of order smaller than $[P_1(\sigma)]^3$:
    We can see that considering the term $P_{S_2}$ in Eq.~(\ref{general sum}) could make all the first and second order terms of $\log[1-P_f(\sigma)]$ vanish, which means $[P_1(\sigma)]^3$ dominates when $\sigma$ is small enough and the slope of cdf of fishnet failure in Weibull scale can increase by a factor of 3.
    
    
    
    \subsubsection{Upper and Lower Bounds of Failure Probability of Fishnet}
    As is already seen in previous sections, the fishnet possesses two geometrical limits i.e. chain (weakest link) and fiber bundle: the structure reduces to a chain if we set the aspect ratio to $m:n=1:N$ and to a fiber bundle if we set the ratio to $m:n=N:1$ (Loads are applied horizontally). Intuitively, given the same amount of links, a fiber bundle would be much more reliable than a single chain: failure of one link will not necessarily lead to rupture of the whole bundle. Therefore, we conjecture that the upper and low bounds of fishnet's failure probability should be that of a weakest link and fiber bundle consisting of the same number ($N$) of links, respectively (see Fig.~\ref{fig:Geometric Transform}). Furthermore, by assuming that $P_f(\sigma)$ changes continuously as the fishnet gradually transforms from a chain to a bundle, the failure probability decreases when the aspect ratio $m/n$ increases at a fixed value of the product $m\times n = N$ .
    \begin{figure}[!h]
        \centering
       \includegraphics[width=0.6\textwidth]{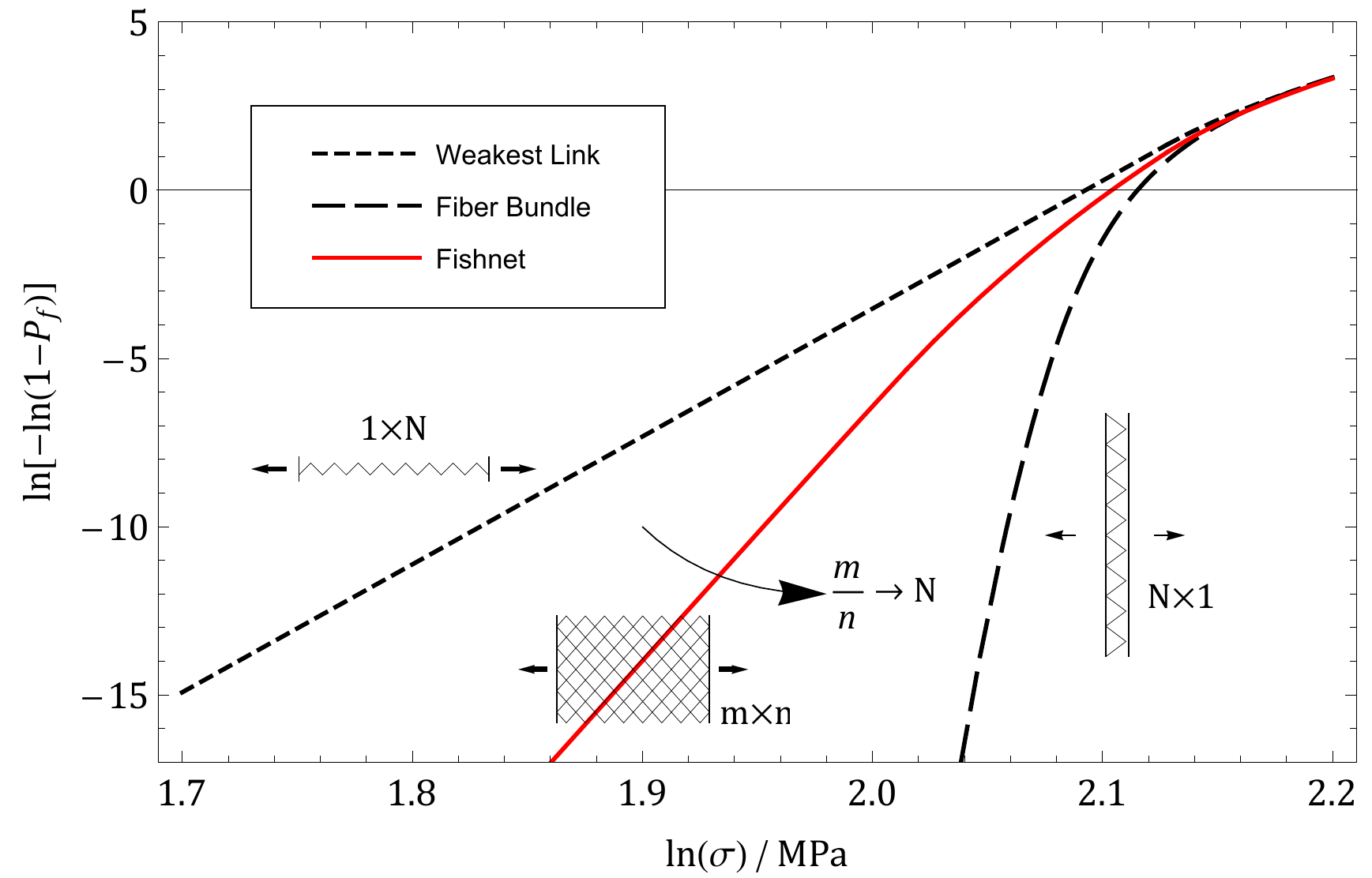}
       \caption{Schematic showing the change of failure probability of fishnet by changing its aspect ratio $m/n$ gradually from $1:N$ to $N:1$.}
       \label{fig:Geometric Transform}
    \end{figure}
    
\section{\large Monte Carlo Simulation of Fishnet Failure}


\subsection{Setup of Finite Element Model}
    
    To verify the theory, numerical Monte Carlo experiments are conducted by finite element simulations (in Matlab). In the finite element model (FEM) model, the size of rectangular fishnet is defined by its number of rows and columns $m \times n$. Each link is treated as an elastic truss element. All the elements (or links) have the same cross section area and Young's modulus, and differ only in their strength, $s_i$, which is a random variable that follows the probability distribution $P_1(\sigma)$. The nodes at the left fishnet boundary are fixed in $x$ direction but allowed to slide freely in the transverse $y$ direction. The fishnet is loaded under displacement control. The tensile load is applied by prescribing increments of a uniform displacement $u_0$ at the right boundary nodes while allowing transverse free sliding. 
    
\subsection{Random Variable Generation}

Random variables $s_i$ are generated by the inverse-transform method. First, we use the random number generator in Matlab to generate random variables $x_i$ in the interval $[0,1]$, and then apply the inverse of $P_1$ to each $x_i$, i.e. $s_i=P_1^{-1}(x_i)$. Based on previous research\cite{BazLeBaz09}, $P_1(s)$ is defined by grafting a power-law left tail onto a Gaussian distribution. 
   
The stiffnesses of links ($EA/l_e$) are not treated here as random variables. This is not a major simplification. To explain why not, consider an $m\times n$ fishnet with i.i.d. random stiffness $K_{i,j}$ for each link. Now, if the $x$-displacement at left boundary is zero and on the right boundary is $n u_0$, then the elongation of each link is $u_0$. So, if we pick any cross-section, the total force and nominal stress are:
    \begin{equation}
    \label{LLN}
        P=u_0\sum_{r=1}^{m}K_r,~~~\sigma_N=\frac{u_0}{A}\left( \frac{1}{m}\sum_{r=1}^{m}K_r \right)
    \end{equation}
where $K_r$ are the element stiffnesses along the $r$-th cross-section of fishnet, and $A$ is the cross-section area of each element. This means that the random variable $\sigma_N$ is proportional to the sample mean of $K_r$. According to the weak law of large numbers 
the sample mean converges to true mean with probability 1 as $m\rightarrow \infty$. Intuitively, this means that given any form of the distribution of $K_{i,j}$, the nominal stress $\sigma_N$ always follows the law of a degenerate Gaussian distribution, i.e., the Dirac Delta function, centered at $\sigma_N=Ku_0/A$ if the cross-section contains a sufficiently large number of elements, where $K$ is the mean of $K_r$. Therefore, no matter what the distribution is for the stiffness of a single element, its influence on the randomness of nominal strength negligible. This is also true if the fishnet is slightly damaged, because we can always find a cross-section that is sufficiently far away from the zone of failed links. Indeed, the elongation of these elements on the chosen cross-section is still approximately $u_0$, which means that Eq.~(\ref{LLN}) still holds. Therefore, it suffices to randomize strength of each link while considering the element length, elastic modulus and cross-section area as deterministic.

\subsection{Element Deletion and Propagation of Cracks}

    The time steps are indexed by the total number ($k$) of failed links. At the beginning of time step $k$, we set $u_0=1$ and calculate the stresses of each element $\sigma_j^{(k)}$. Then the only element, indexed by $i$, that is going to fail in this new time step is the one whose stress satisfies the condition:
    \begin{equation}
     \lambda^{(k)}=s_i/\sigma_i^{(k)}=\min_j\ \{s_j/\sigma_j^{(k)} \}
    \end{equation}
where $s_j$ is the strength of the $j^{th}$ element. Since the constitutive law is linear elastic, $\lambda^{(k)}$ is the load multiplier such that if we reset $u_0 = \lambda^{(k)}$ and recalculate the stress field, we will have $\sigma_i^{(k)} = s_i$ and $\sigma_j^{(k)} < s_j$~for~$i\neq j$. So the final stress state of each element at the current time step is calculated by $\sigma_{j_{\text{new}}}^{(k)} = \lambda \ \sigma_i^{(k)}$. After updating the stresses, the critical $i^{th}$ element is deleted, and its element stiffness matrix $\bm{K}_{e^i}^{(k)}$ will not be assembled into the global stiffness matrix $\bm{K}^{(k)}$ for future loading steps. This process keeps going until the global stiffness matrix becomes singular, which means that a crack has already gone through the cross-section and the fishnet has failed.
    
Indexing the time steps by the number of failed links allows us to obtain not only the peak load but also the complete load-displacement curve without worrying about stability issues. In fact, the load-displacement curve of elastic fishnet shows a strong snap-back instability, as discussed paper.

\subsection{Results of FE Simulation}

    Fig.~\ref{fig:FE Simulation}(a) shows the load-displacement curve of a typical result of FE simulation on $16 \times 32$ fishnets with random strength. The distribution of random strength is a Gaussian distribution $N\sim(10,0.4^2)$ grafted with a power law tail at $P_1(\sigma)=0.015$. Often, peaking of the load and a strong snap-back instability is observed right already before the rupture of the first or second link. This is no surprise because, if there is no randomness in strength, the fishnet will reach the maximum load (and fail) as soon as any one of the links succumbs. 

So it is the strength randomness that makes it possible for the whole structure to survive after failure of a few links. On the other hand, in real nacreous materials, in which the shear connections between the lamellae do not act locally as pin nodes and resist bending moments, a few links (or shear connections) will probably have to break before the maximum load is reached. This would be similar type I failure in brittle heterogenous materials, in which several microcracks between grains must form in a representative volume element (RVE) before a macrocrack propagation from the RVE is initiated. Thus it might be necessary to generalize the present theory by stipulating that a certain small number of links forming an RVE must break before the maximum load is reached. This means that the true nacreous behavior would ba a transition from Type I fracture to a fishnet fracture. Such a generalization, however, must be relegated to further work.

    \begin{figure}[!h]
        \centering
       \includegraphics[width=0.98\textwidth]{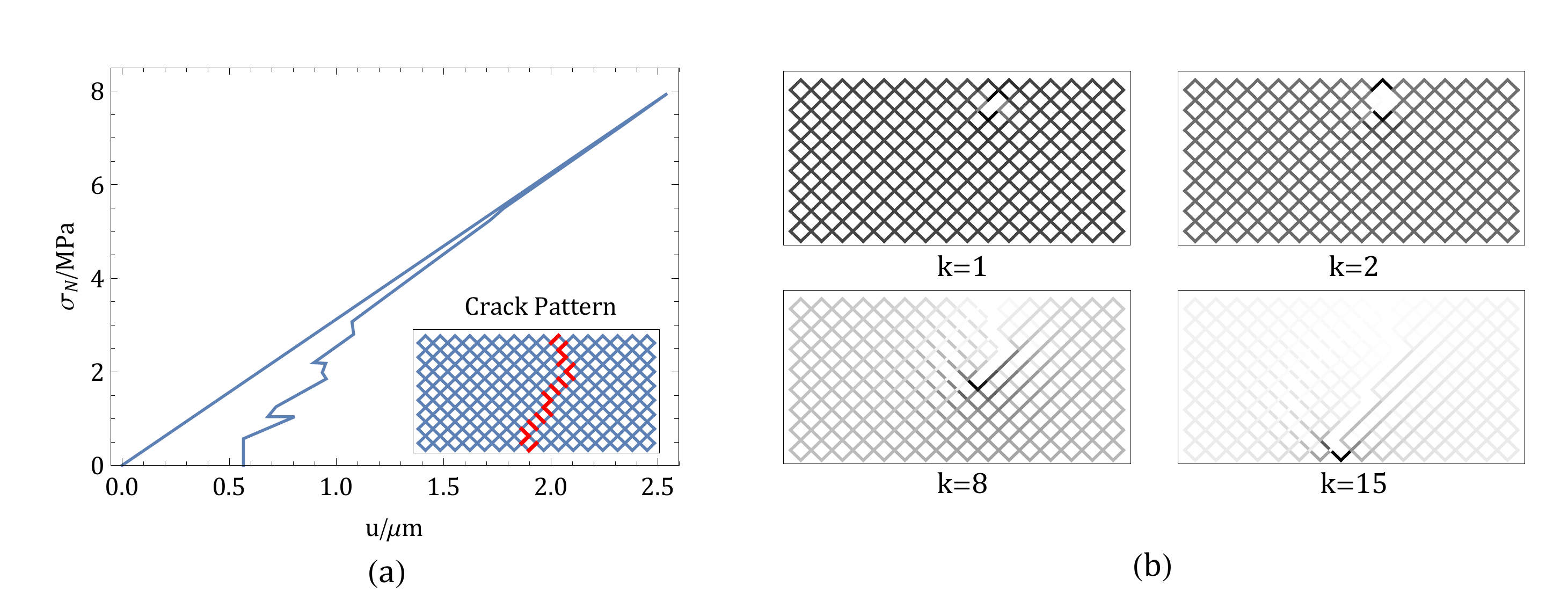}
        \caption{a) Load-displacement curve of a $16\times32$ fishnet with random strength together with its crack pattern; (b) Stress evolution as crack propagates through the fishnet (the darker the color, the larger the normalized stress $\sigma/\sigma_{max}$), where $k$ is the number of failed links.}
        \label{fig:FE Simulation}
    \end{figure}
    
Fig.~\ref{fig:FE Simulation}(b) illustrates the stress field of 4 typical fishnet frames ($k=1,~2,~8,~15$) in the process of crack propagation. In this particular case, subsequent failures localize immediately after the crack initialization, and no scattered damage is observed. One can see that the crack path in this case has a strong tendency to move vertically and form a cross-section, even though the crack path is random. Another observation is that as crack propagates, sharper contrast of darkness in the figures are seen, meaning that the difference between maximum stress $\sigma_{max}$ and nominal stress $\sigma_N$ becomes larger and larger, making the whole structure even weaker.

    \subsection{Verification of Fishnet Theory}
    
Millions of finite element simulations are run to verify the fishnet statistics. The criterion of convergence is set as follows: First we consider the curve of the estimated distribution obtained from many runs of fishnet sample of size $n_1$ in Weibull scale. Then we increase the sample size $n_1$ by one and consider the curve of the newly estimated distribution. The part of the curve that does not change much upon doubling the sample size is considered as converged. 

To obtain an accurate estimate of the distribution, especially in its left tail, at least on million ($10^6$) samples have been computed for each case. Based on this scenario, the distribution estimated from the histogram of the results of all computer runs converges very well to the exact distribution in the region $Y = \ln[-\ln(1-P_f)] > - 10$ in Weibull scale. This corresponds to the interval $4.54\times10^{-5} \leq P_f <1$.

\subsubsection{Two-Term Fishnet Model}
    \begin{figure}[!h]
        \centering
        \includegraphics[width=1\textwidth]{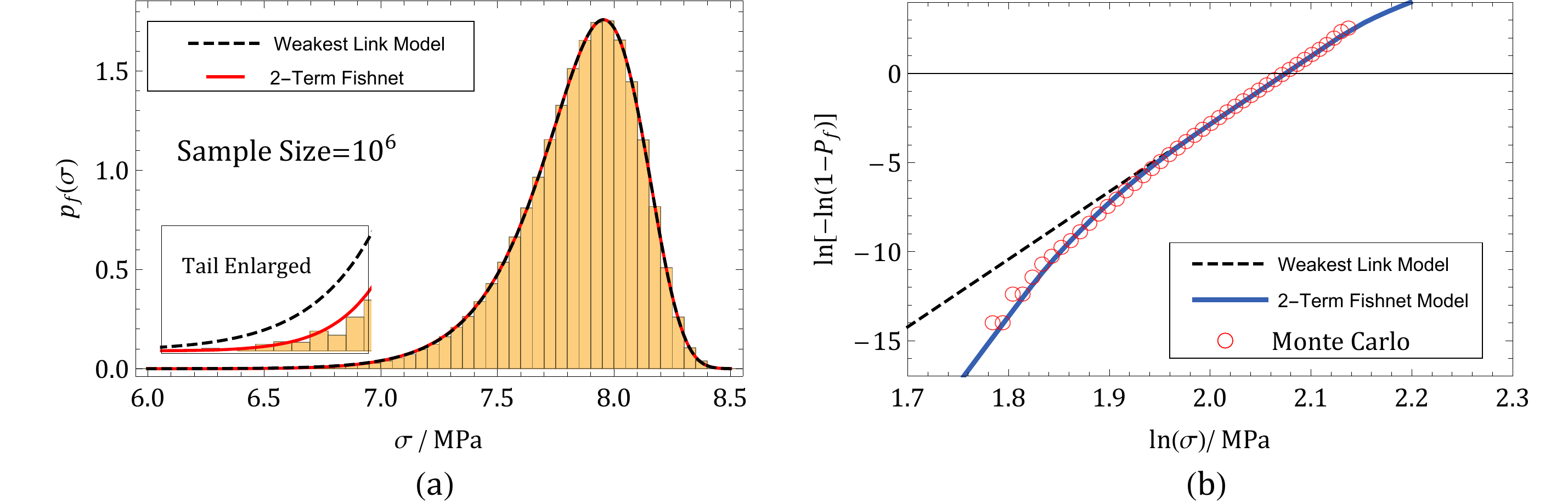}
        \caption{(a) Histogram of Monte Carlo Experiment of $10^6$ samples (frequency is normalized to probability density) on $16 \times 32$ fishnets compared with the density functions of weakest link model and 2-term fishnet model; (b) Same data converted into cumulative probability and plotted in Weibull scale.}
        \label{fig:histogramPG}
    \end{figure}
    When $P_1(\sigma)$ is a Gaussian $N(10,0.8^2)$ grafted with a power law tail at $P_f=0.015$, its tail is relatively light and 2-term fishnet model works very well. The expression of $P_1(\sigma)$ is chosen as
    \begin{equation}
    \label{P1 PG}
        P_1(\sigma) = 
       \begin{cases}
           0.11338\cdot (\sigma/10)^{38}, & \sigma \leq 8.4~MPa\\
           0.015 + \left\{0.481 - 0.504\cdot\text{Erf}[0.884 (10 - x)]\right\}, & \sigma>8.4~MPa
       \end{cases}
    \end{equation}
    where
    \begin{equation}
       \text{Erf}(x)=\frac{2}{\sqrt{\pi}}\int_{x}^{\infty}e^{-t^2}dt.
    \end{equation}
Fig.~\ref{fig:histogramPG}(a) shows the estimated probability density function (pdf), $p_f(\sigma) = \dd P_f(\sigma)/ \dd \sig$, obtained from the histogram of Monte Carlo simulations, and compares this pdf with the prediction of the weakest-link model (black dashed line) and of the two-term fishnet model (red continuous line). These two predictions are almost indistinguishable in the range $\sigma>6.7 MPa$ and match very well the Monte Carlo experiments. However the difference shows up in the tail ($\sigma \leq 6.7 MPa$). Indeed, the two-term fishnet model fits the Monte Carlo data much better than the classical weakest-link model. The difference is more obvious when it is plotted in Weibull scale (Fig~\ref{fig:histogramPG}(b)): the results of Monte Carlo simulations match the two-term fishnet model everywhere in the figure, while the weakest link model works well only in the range $\ln\sigma>1.9$.

\subsubsection{Three-Term Fishnet Model}
    
The effect of $P_{S_2}$ in Eq.~(\ref{general sum}) becomes more significant if the tail of $P_1(\sigma)$ gets heavier. Now consider the case where $P_1(\sigma)$ is a grafted distribution of Gaussian $N(10,0.8^2)$ and a power law tail with the grafting point $P_f(\sigma)=8.955\%$: 
    \begin{equation}
    \label{P1 WG}
        P_1(\sigma) = 
      \begin{cases}
           2.551 \cdot \left[ 1-e^{-(\sigma/12)^{10}} \right], &
            \sigma \leq 8.6~MPa\\
           0.08955 + \left\{ 0.436 - 0.474\cdot \text{Erf}[0.884 (10
            - x)] \right\}, & \sigma>8.6~MPa
   \end{cases}
    \end{equation}

To see the effect on the left tail "thickness", we calculate that, for $\sigma=6.7 MPa$, a change of the grafting point from $P_1(\sigma) = 0.1\%$ to $P_1(\sigma) = 9\%$ increases $P_1$ from $2.79\times10^{-6}$ to $7.5\times10^{-3}$ (note that these grafting point values are much higher than what was identified for particulate composites \cite{BazLe17}). This increase makes the tail much "thicker" than in the previous case. Once again, for this particular strength distribution $P_1(\sigma)$ of a single element, and for the $16 \times 32$ fishnet, one million samples of Monte Carlo simulations have been obtained.

    \begin{figure}[!h]
        \centering
       \includegraphics[width=1\textwidth]{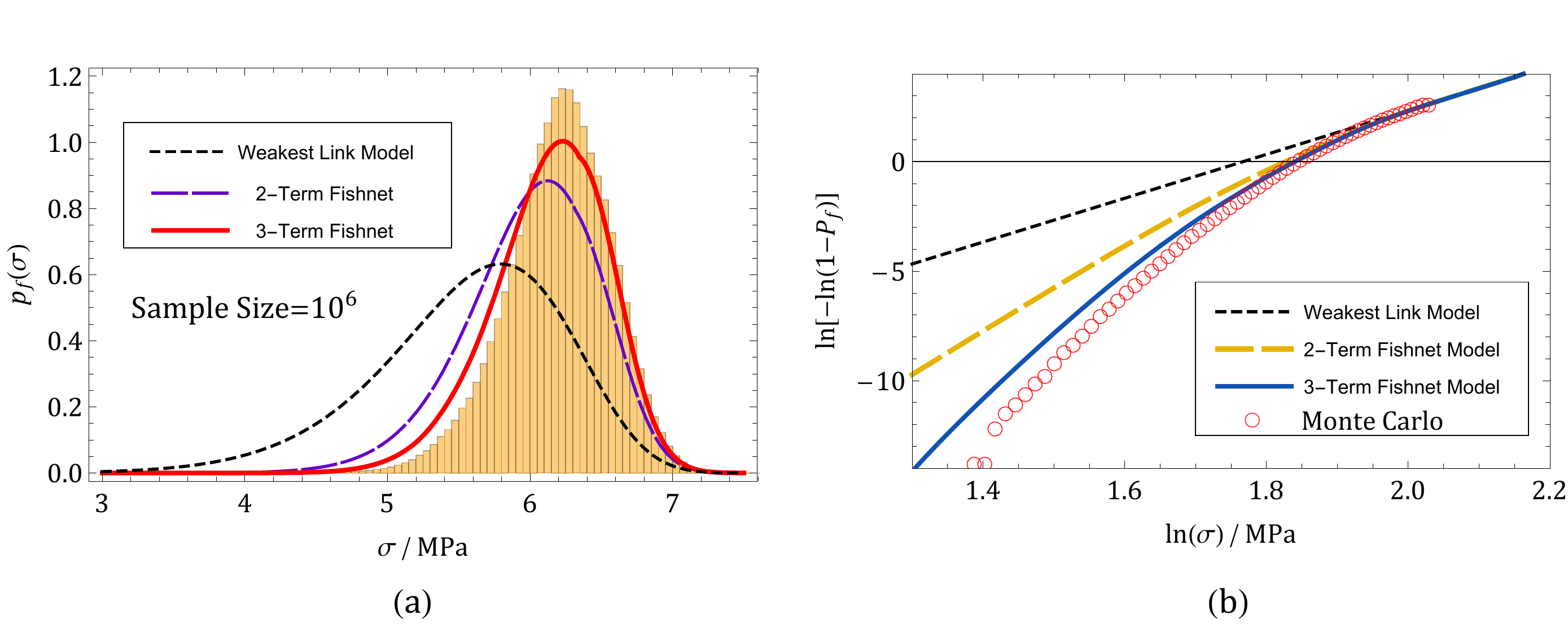}
        \caption{(a) Histogram of Monte Carlo Experiment of $10^6$ samples (frequency is normalized to probability density) compared with the probability density functions of weakest link model, 2-term and 3-term fishnet model; (b) Same data converted into cumulative probability and plotted in Weibull scale.}
        \label{fig:histogramWG}
    \end{figure}    

Fig.~\ref{fig:histogramWG} shows the estimated density function of $P_f$ with $P_1(\sigma)$ as mentioned above and its comparison with the predictions of the weakest link model, 2-term (Eq.~(\ref{extension})) and 3-term (Eq.~(\ref{3-term fishnet})) fishnet model. We can see that the weakest link model gives a poor prediction of failure probability, while the fishnet models are much better. Note that, as more and more terms of $P_{S_k}$ are considered in the fishnet model, the skewness of the corresponding distribution becomes smaller and smaller. From Fig.~\ref{fig:histogramWG}(b), all of the three predictions are upper bounds of the true failure probability. As $\sigma$ decreases, the Monte Carlo results begin to deviate from the two-term fishnet model and a further slope increase is observed. As we see, the three-term fishnet model gives the best prediction. To get a still more accurate prediction, we would need to consider the next term, $P_{S_3}$, or even more terms, in the expansion of Eq.~(\ref{general sum}).

\subsubsection{Scattered Damage vs. Localized Damage}

The previous Monte Carlo simulations show that changing the tail of strength distribution of a single link could have a huge effect on the failure probability of the whole structure. To explain it intuitively, we run finite element simulations for a $32\times32$ square fishnet for two different strength distributions of a single link $P_1(\sigma)$ and $P_1^*(\sigma)$, where $P_1(\sigma)$ is the Gaussian distribution grafted with a light power law tail (Eq.~(\ref{P1 PG})) and $P_1^*(\sigma)$ is the one with a heavy Weibull tail (Eq.~(\ref{P1 WG})).

    \begin{figure}[!h]
        \centering
       \includegraphics[width=0.95\textwidth]{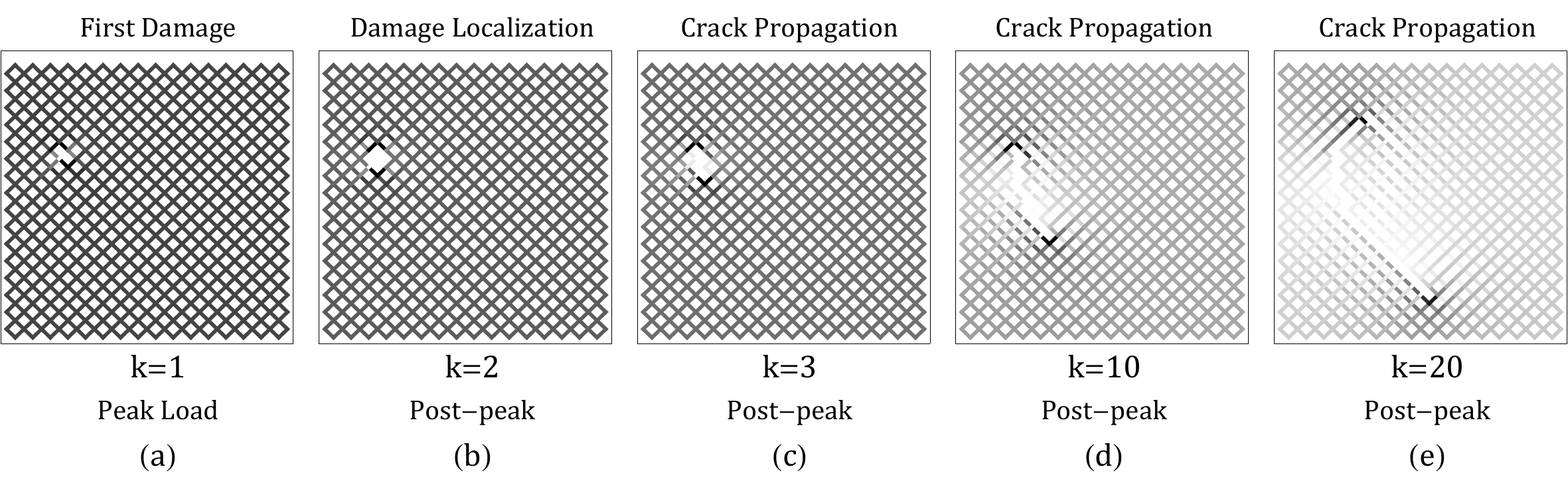}
        \caption{Stress evolution of a $32\times32$ square fishnet whose strength distribution of a single link is $P_1$, which has a thin power law tail. In each figure, the stress is normalize by the maximum $\sigma_{max}^k$ of that moment. Pure black corresponds to $\sigma_{max}^k$ and pure white corresponds to zero stress.}
        \label{fig:stress evolution heavy PG}
    \end{figure}

    First, consider the case of $P_1(\sigma)$. A typical simulation result is shown in Fig.~\ref{fig:stress evolution heavy PG}$(a)\sim (e)$. Due to the fact that $P_1(\sigma)$ has a very thin power law tail, the probability that there exists an extremely weak link is vanishingly small. Therefore, in most cases, the peak load is reached  right before the first rupture. Then the successive failures localize and form a single crack. This means that, with a very high probability, the failure of structure is the result of rupture of a single link. Then the weakest link model predicts the $P_f$ quie accurately, as long as $\sigma$ is not too small.
    
    \begin{figure}[!h]
        \centering
        \includegraphics[width=0.95\textwidth]{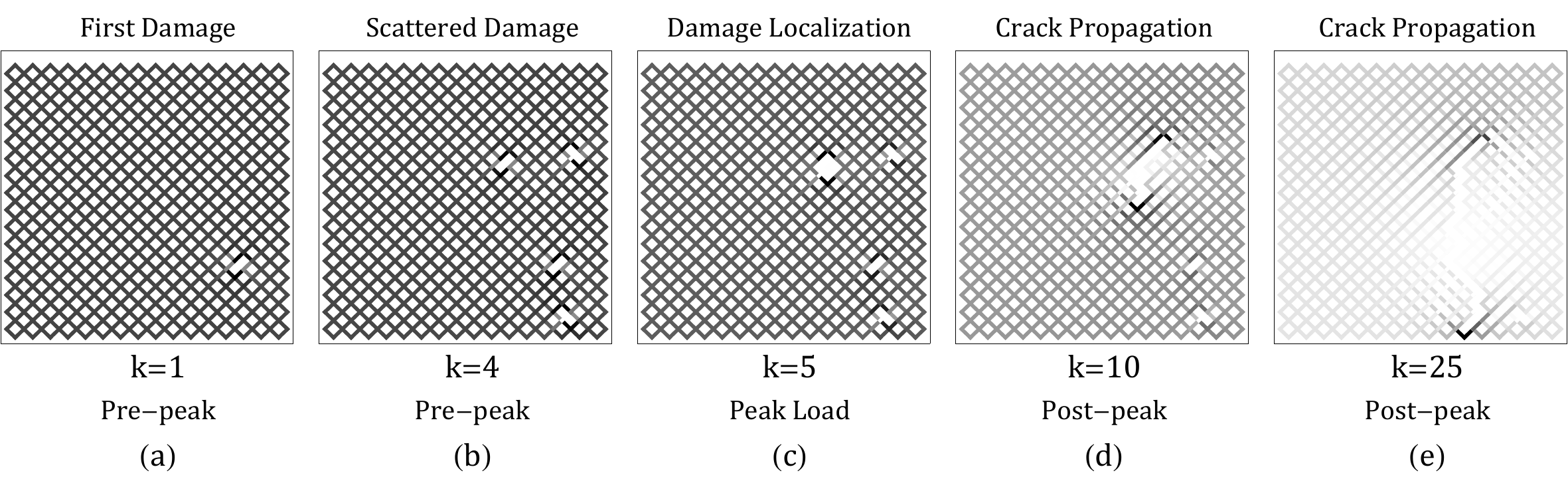}
        \caption{Stress evolution of a horizontally pulled $32\times32$ square fishnet in which the strength distribution, $P_1^*$, of a single link has a thick Weibull tail. In each figure, the stress is normalized by the maximum $\sigma_{max}^k$ reached at that moment. Perfect black corresponds to $\sigma_{max}^k$ and perfect white corresponds to zero stress.}
        \label{fig:stress evolution heavy WG}
    \end{figure}

The scenario gets more complicated as the tail of $P_1$ becomes thicker. Very likely, a few fishnet links have a very low strength and their failures will affect the fishnet strength very little. Moreover, these very weak links are scattered through the whole fishnet. As the loading of fishnet begins, these weakest links will fail successively at very small nominal stresses, while the rest of the "normal" links are still sound and safe. As long as the first few damages are scattered, the stress field will keep to be almost uniform except near these failed links. In this case, the failure of one or two links need not lead to the failure of the whole structure, and the weakest-link model two-term fishnet models cannot give good estimates of $P_f$. This is exactly why we need to consider more higher-order terms in the formula (Eq.~(\ref{general sum})) of survival probability.
    
Fig.~\ref{fig:stress evolution heavy WG} (a)--(e), shows the typical pattern of stress evolution when the strength distribution of links, $P_1^*(\sigma)$, has a thicker Weibull tail. In this particular case, the damage zone does not localize until the fifth rupture, ossurring just after the peak load is reached. Once the damages start to localize, the stress concentration at the crack front will only get higher and higher, making the rest of the links bear smaller and smaller stresses and next failures more likely to localize. Consequently, the nominal stress keeps decreasing. In this process, damage localization offers a positive feedback to the system: the current localization leads to a higher stress concentration, making the next failure more likely to localize. Such a process keeps going until the structure fails. Eventually the stability of fishnet (under load control) is lost and the peak load occurs right at the moment of damage localization, which depends strongly on the "heaviness" of the tail of $P_1(\sigma)$.

\subsubsection{Transition from Chain to Bundle (Shape Effect)}

Based on the fact that the peak load is most likely reached at the onset of damage localization helps to decide how many terms of $P_{S_k}$ are needed to get an accurate enough prediction of $P_f$: For a given $P_1(\sigma)$ of strength distribution of links and a given size and shape of fishnet, the approximate number of terms we need to add in the fishnet model should be greater than the average number of scattered failures before localization, and the more terms get added, the more accurate the estimation will be in the tail of $P_f$.

It thus becomes clear that the fiber bundle must give the lower bound on the strength of all fishnets consisting of the same number of links. The fiber bundle has the largest cross section, and no stress concentration. So the stability limit of a bundle is reached the latest and each term $P_{S_k}$ is the largest among all fishnets. Therefore we have to consider most terms $P_{S_k}$ in the expansion of survival probability Eq.~(\ref{general sum}) so as to get an accurate estimation of $P_f$. Hence, the survival probability, $1-P_f$, of a bundle is the upper bound for those of all fishnets and, accordingly, the failure probability, $P_f$, of a bundle is the lower bound on all the failure probabilities of all fishnets.
    
   \begin{figure}[!h]
       \centering      
       \includegraphics[width=0.95\textwidth]{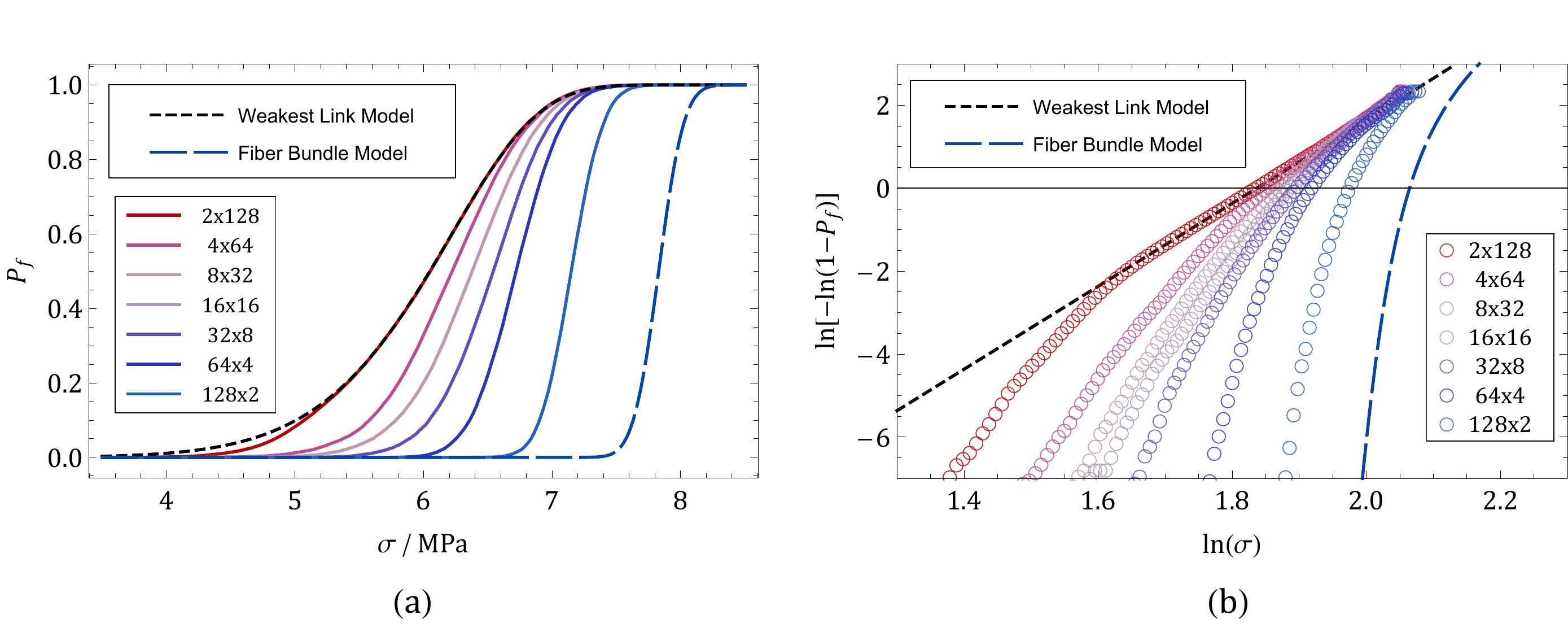}
        \caption{a) Monte Carlo simulations showing the transition of $P_f$ as aspect ratio of fishnet is changed from $1\times N$ to $N \times 1$; b) Same data plotted in Weibull scale.}
        \label{fig:transition fishnet}
    \end{figure}

Fig.~\ref{fig:transition fishnet} shows the transition of failure probability $P_f$ of fishnets of various aspect ratios, consisting of the same number, $N$, of links. The circles with different colors corresponds to results of Monte Carlo simulations. 
Since we want to verify only the qualitative effect of aspect ratio on $P_f$, we reduce the sample size from $10^6$ to $2\times10^5$. Accordingly, we focus only on the range  $\ln[-\ln(1-P_f)] \geq -6$, in which the histogram has converged quite close to the true distribution.
    
As shown in the figure, the   
weakest link model (black dashed line) is the strict upper bound of all fishnets with the same number of links, and the fiber bundle (dark blue dashed line) 
is the strict lower bound. When the fishnet is long and thin, for example $m \times n = 2 \times 128$ (dark red circles), the corresponding $P_f$ can be described quite well by the two-term Fishnet model, which shows a gradual slope increase by a factor of 2 as $\sigma_N = \sigma$ tends to 0. 

This last case might look similar to the real nacre shell, which is much longer than wide. However, in nacre shells and similar structures, the highly stressed zone of often highly localized, as in bending, and then the shell may rather behave as our fishnet with a similar length and width (the statistics of fishnets under flexure, or transverse stress gradient, as well as the statistics of three-dimensional fishnet generalization, would probably be similar, but its study must be relegated to a subsequent article).  

    
Returning to the uniformly stressed long and narrow fishnet, it cannot be described adequately by the two-term fishnet, because this model applies if only one additional link fails before the maximum load.
The higher-order terms in the expansion of $P_f$ matter. Although they are hard to calculate analytically, insight can be gained by Monte Carlo simulations. We run simulations for fishnets of size $64 \times 16$ and $16 \times 64$, and then count the average number $r_p$ of scattered link failures before the peak load is reached (the distribution of $r_p$ should in fact lead to the Poisson distribution with rate parameter $\mu_p$ when total number of links $N\rightarrow \infty$). For fishnets of size $64 \times 16$, $\mu_p \simeq 5.2$, while $\mu_p \simeq 4.4$ for fishnets of size $16 \times 64$. This shows that more terms of $P_{S_k}$ are needed for short and wide fishnets than for the long and narrow ones. Due to the fact that the laws of stress redistribution around scattered failures are almost identical, the common terms of $P_{S_k}, k \leq k_0$ in $P_f$ for both fishnets are almost the same. The only difference is that the next few terms $P_{S_k}, k > k_0$ for the short and wide fishnets are much larger than those for the long and narrow ones. The cause is the delay of damage localization.


\subsection{Comparison with the Chain of Bundles}

Attempts have previously been made \cite[e.g.]{Har78I,Har78II,Wei15} to use a chain of fiber bundles to model the failure probability of fiber composites under uniaxial tension. A long specimen under uniaxial tension is subdivided, by imagined cross sections of an assumed spacing, into fictitious segments, each of which is modeled as a fiber bundle. This is an approach that has physical basis in the microstructure of a fiber composite but not of nacre-like materials.

If a fiber bundle with equal load sharing is represented mechanically by loading through rigid platens, each parallel coupling reduces the reach of a power law tail by about one order of magnitude \cite{BazPan06, BazLe17}. For 10 parallel fibers, the distribution becomes Gaussian except for a power law (or Weibull) tail reaching to the probability of only about $10^{-20}$. The parallel couupling of fibers extends the reach of the Weibull tail \cite{BazPan06, BazLe17}, but about $10^{22}$ bundles in the chain would be needed to approach the Weibull distribution expected for a very long structure. Obviously, such mechanics based equal load-sharing hypothesis gives clearly unreasonable predictions. 

The problem with power law (or Weibull) tail shortening is avoided by some convenient, non-mechanical, purely intuitive, load-sharing rules for the transfer of load from a failed fiber to its neighbors. A convenient load-sharing rule can give a very different probability distribution $G_n$ for each bundle, with a realistic reach of power law tail. Like the fishnet, a chain of bundles, each with a suitable $G_n$, can then predict, for low probabilities, a reasonable slope increase in the low-probability range of Weibull plot, as shown by Harlow and Phoenix \cite{Har78I,Har78II}.

In the fishnet model, by contrast, this slope decrease is due to the addition of non-zero probability $P_{S_k}$ of structure survival after a failure of $k$ links. The slope of $P_f$ will increase, at least by a factor of 2, due to adding the first term, $P_{S_1}$. As mentioned before, adding more terms $P_{S_k}$ increases the slope further. Importantly, no separate hypothesis about transverse load sharing or redistribution is needed in the fishnet model to get the above result.

\section{\large Conclusions}
               \begin{itemize} \setlength{\itemsep}{-1.5mm}

        \item  Based on similar mechanical responses, nacre-like imbricated laminar materials under uniaxial tension can be modelled as fishnets with square holes pulled in the diagonal direction.
        \item  Severe stress redistribution around isolated damages of fishnets are confined in their close neighborhood, but greatly affect the failure probability. 
        \item  Assuming the link strength values in a $m \times n$ fishnet to be i.i.d. random variables characterized by failure probability $P_1(\sigma)$, one can express the fishnet failure probability as a finite series $P_f(\sigma)= 1 - \sum_{k=0}^{m(n-1)} P_{S_k} (\sigma)$ in which $P_{S_k}(\sigma)$ is the probability of the event that total of $k$ links have failed while the structure is still safe, while $P_{S_0}$ corresponds to the classical survival probability of a chain with $m\cdot n$ links. Apart from the first term in the series, the remaining terms tend to 0 as $\sigma$ approaches $\infty$ and tend to $c_k [P_1(\sigma)]^k$ as $\sigma$ approaches to 0. Consequently, in the upper tail $P_f$ is essentially the same as the tail of weakest link model while in the lower tail, $P_f$ exhibits a major deviation from the weakest link model and curves downward in the Weibull scale.
        \item As $\sigma$ decreases, the slope of $P_f$ in the Weibull plot gets steeper by a factor of 2 due to the effect of $P_{S_1}$. The center of transition to increased slope depends on the level of stress concentration near the failed link. A higher stress concentration causes the transition to happen at smaller $\sigma$. The further term in the expansion cause smaller and smaller increases of slope of $P_f$ in Weibull plot, and the subsequent slope transitions are centered at smaller $\sigma$ values, due increasing stress concentration.
        \item The fishnet probability $P_f$ exhibits strong shape effects. By changing the aspect ratio $m/n$ from $1/N$ to $N/1$, a fishnet gradually transforms from a chain to a fiber bundle, and the corresponding failure probability transits smoothly from the Weibull distribution to the Gaussian (or normal) distribution. When the shape $m\cdot n$ is fixed, a larger value of $m/n$ corresponds to a strictly smaller $P_f$ for all $\sigma$. So, when $m \cdot n$ is fixed, the weakest-link model and fiber bundle model give, respectively, the strict upper and lower bounds of the failure probability, $P_f$. 
        \item There is a strong size effect, similar to, though different from, the Type 1 quasibrittle size effect characterizing particulate or granular materials and fiber composites.
    \end{itemize}

\listoffigures

\end{document}